\documentclass[usenatbib,useAMS]{mn2e}

\usepackage{amsfonts}
\usepackage{amsmath}
\usepackage{wasysym}
\usepackage{graphicx}
\usepackage{longtable}
 
\begin{document}

\label{firstpage}

\title[Internal dynamics of early type galaxies]{Understanding the internal dynamics of elliptical galaxies without non-baryonic dark matter}

\author[Dabringhausen et al.]{
J. Dabringhausen$^{1}$ \thanks{E-mail: joerg@astro-udec.cl},
P. Kroupa$^{2,3}$ \thanks{pavel@astro.uni-bonn.de},
B. Famaey$^{4}$ \thanks{benoit.famaey@unistra.fr} and
M. Fellhauer$^{1}$\thanks{mfellhauer@astro-udec.cl}\\
$^{1}$ Departamento de Astronom\'{i}a, Universidad de Concepci\'{o}n, Casilla 160-C, Concepci\'{o}n, Chile\\
$^{2}$ Helmholtz-Institut f\"{u}r Strahlen- und Kernphysik, Universit\"at Bonn, Nussallee 14-16, 53115 Bonn, Germany\\
$^{3}$ Charles University in Prague, Faculty of Mathematics and Physics, Astronomical Institute, V  Hole\v{s}ovi\v{c}k\'ach 2,\\
 CZ-180 00 Praha 8, Czech Republic\\ 
$^{4}$ Observatoire astronomique de Strasbourg, Universit\'e de Strasbourg, CNRS UMR 7550, 11 rue de l'Universit\'e, 67000 Strasbourg,\\
 France}

\pagerange{\pageref{firstpage}--\pageref{lastpage}} \pubyear{2016}

\maketitle

\begin{abstract}
Assuming virial equilibrium and Newtonian dynamics, low-mass early-type galaxies have larger velocity dispersions than expected from the amount of baryons they contain. The conventional interpretation of this finding is that their dynamics is dominated by non-baryonic matter. However, there is also strong evidence that many low-mass early-type galaxies formed as tidal dwarf galaxies, which would contain almost no dark matter. Using an extensive catalogue of early-type galaxies, we therefore discuss how the internal dynamics of early-type galaxies in general can be understood by replacing the assumption of non-baryonic dark matter with two alternative assumptions. The first assumption is that Milgromian dynamics (i.e., MOND) is valid, which changes the effective gravitational force in the weak-field limit. The second assumption is that binary stars affect the observed line-of-sight velocity dispersions. Some moderate discrepancies between observed and predicted velocity dispersions remain also when these effects are implemented. Nevertheless, the observed velocity dispersions in early-type galaxies can then easily be explained without invoking the presence of non-baryonic dark matter in them, but with already documented variations of the galaxy-wide stellar initial mass function and non-equilibrium dynamics in some of the low-mass early-type galaxies.
\end{abstract}

\begin{keywords}
galaxies: dwarf -- galaxies: elliptical and lenticular, CD -- galaxies: kinematics and dynamics
\end{keywords}

\section[Introduction]{Introduction}
\label{sec:introduction}

The currently prevailing cosmological model is the $\Lambda$CDM-model, which is named for the two principal components of the energy content of the Universe according to that model. These components are dark energy (represented by a cosmological constant $\Lambda$ in Einsteins' field equations) and non-baryonic cold dark matter (CDM).

According to that model, the Universe has expanded to its present-day volume from an extremely dense configuration, and the initial distribution of CDM was not completely homogeneous. These inhomogeneities grow with time, and this effect is the strongest in the regions with the highest initial densities. Thus, the density contrast between the densest regions of the Universe and the overall density of the Universe increases with time. When the density contrast between a certain region of space and the overall density of the Universe surpasses a critical value, this region decouples from the overall expansion of the Universe completely, and the CDM within it collapses into a self-gravitating CDM-halo.

Out of the baryons that are captured within the CDM-haloes during their collapse, the first galaxies of the universe are thought to form. The number of CDM-haloes predicted by the model decreases with increasing mass \citep{Guo2011}, so that the $\Lambda$CDM-model predicts a large number of primordial dwarf galaxies (PDGs) with masses $M<10^{10}\, {\rm M}_{\odot}$ that form within CDM-haloes. 

On the other hand, in encounters between gas-rich disk galaxies, the tidal forces acting on the disks can lead to the formation of long tidal tails in which self-gravitating structures may form, as numerical calculations have shown \citep{Barnes1992a,Elmegreen1993,Bournaud2010}. These newly formed structures have masses of up to $10^9 \ {\rm M}_{\odot}$ \citep{Elmegreen1993,Bournaud2006} and have radii of the order of $\sim$1 kpc \citep{Wetzstein2007,Bournaud2008}. It has also been shown that these objects can survive on a time-scale of $10^9$ years with sites of long-lasting star formation \citep{Bournaud2006,Recchi2007,Ploeckinger2014}. Given these properties, the structures emerging in the tidal tails can be considered galaxies (cf. \citealt{Bournaud2007,Forbes2011}). Due to their small mass and their tidal origin, such galaxies have been named tidal dwarf galaxies (TDGs).

The TDGs consist only of baryonic matter, even if the progenitors of the TDGs contained a substantial amount of CDM \citep{Barnes1992a,Duc2004,Bournaud2006}. The reason is that the baryonic matter that becomes part of a TDG has before the encounter been part of thin rotating disks, and therefore occupied only a small region in phase space. CDM, on the other hand, is distributed spheroidally and is made up of particles on random orbits, so that its density in phase space is low, while it occupies a large volume. However, in an encounter, only matter within the same region of phase space can end up on similar trajectories in phase space, while this is a prerequisite for the formation of a bound object through gravitational collapse \citep{Bournaud2010}. Given that TDGs thus contain only negligible amounts of CDM, in contrast to PDGs with the same amount of baryonic matter, it would be natural if the PDGs and the TDGs would constitute populations that are distinguishable by their properties; not only due to their different origin, but also due to their different composition \citep{Kroupa2010,Kroupa2012,Dabringhausen2013,Kroupa2015}.

Interacting galaxies with bridges of matter connecting them or elongated arc extending from them have indeed been observed (for instance the Antennae Galaxies and the Mice Galaxies), and already \citet{Zwicky1956} argued that these filaments can be explained with tidal forces acting between the galaxies, and that new galaxies may form in them. The observed filaments indeed closely resemble the tidal tails that appear in numerical calculations of galaxy encounters and subsequent TDG-formation (e.g. \citealt{Barnes1992a,Elmegreen1993,Bournaud2006,Wetzstein2007}). Observed stellar systems that are candidates for TDGs due to their age and their vicinity to tidal tails to have been found by numerous authors (e.g. \citealt{Mirabel1992,Duc1994,Duc1998,Monreal2007,Yoshida2008,Duc2011,Duc2014}).

The observed counterparts of the PDGs are usually thought to be dwarf elliptical galaxies (dEs; see \citet{Ferguson1994} and \citet{Lisker2009} for reviews on dEs) and dwarf spheroidal galaxies (dSphs; see \citet{Mateo1998} for a review on dSphs). As far as their radii and ages are concerned, the observed dEs and dSphs have properties that seem consistent with the properties expected for PDGs (see, e.g., \citealt{Li2010} and \citealt{Guo2011}). If the low-mass dEs and dSphs are assumed to be in virial equilibrium and that Newtonian dynamics is valid in them, they also have mass-to-light ratios that cannot be explained with the amount of baryons they contain \citep{Mateo1998,Strigari2008,Wolf2010,Tollerud2012,Toloba2014}. This seems consistent with the notion that these galaxies are embedded in CDM-haloes. To a lesser extent, elevated mass-to-light ratios have also been found for more massive and more luminous dEs \citep{Toloba2014}. Note that according to \citet{Ferguson1994}, there is no fundamental difference between dEs and dSphs except for their mass. For this reason, dSphs are considered low-mass dEs in the following.

Thus, on first sight it seems indeed like two distinct populations of dwarf galaxies can be identified in observations. The old, seemingly dark-matter dominated systems could be identified with the PDGs, and the young, star-forming systems near interacting galaxies can the identified with the TDGs, which supports the $\Lambda$CDM-model \citep{Kroupa2010}. However, while observed TDG-candidates usually are young, TDGs could theoretically also be almost as old as the Universe \citep{Bournaud2006,Ploeckinger2014,Ploeckinger2015}. The low age of most observed TDG-candidates is most likely a selection effect, because they are usually identified by their close vicinity to interacting galaxies. However, as time passes by, the TDGs are not necessarily close to their progenitors any longer, while the indications for an interaction between the parent galaxies become harder to observe. This raises the question what old TDGs would look like, and whether there are two old populations of which one can be associated with the PDGs and the other with the TDGs. 

\citet{Dabringhausen2013} argued that the dEs may be old TDGs, since the young observed TDG-candidates would evolve naturally onto the sequence that dEs constitute in mass-radius parameter space. This notion is supported by the fact that the spatial distribution of dEs in the Local Group is highly anisotropic. Most of them can be attributed to one of four planar structures \citep{Pawlowski2013}. The best known of these planar structures are the disks of satellite galaxies (DoS) around the Milky Way (MW) and the Andromeda Galaxy (M31) \citep{Kroupa2005,Metz2007,Pawlowski2012a,Ibata2013}. These DoSs appear to be rotationally supported structures \citep{Metz2008,Ibata2013,Hammer2013}, and evidence for rotation being a common feature of DoSs has also been found outside the Local Group \citep{Ibata2014}. The formation of such rotationally supported planar structures would be natural if the low-mass dEs that constitute them are TDGs, but difficult to understand if they formed as primordial structures in agreement with the $\Lambda$CDM-model \citep{Kroupa2005,Pawlowski2012b,Pawlowski2014}. Some specific models for the formation of the satellite systems of the Milky Way and M31 via the creation of TDGs have been tested successfully in numerical simulations \citep{Fouquet2012,Hammer2013,Yang2014}. Also, the number of satellite galaxies tends to be higher the more pronounced the bulge of a galaxy is, which again supports the notion that satellite galaxies are created through interactions and merger of primordial galaxies, since also the host galaxies become dynamically hotter through interactions \citep{Kroupa2010,LopezCorredoira2016}. Thus, there is evidence for the dEs being old TDGs instead of PDGs. However, if this is correct, the high mass-to-light ratios derived for many dEs under the assumptions of virial equilibrium and Newtonian dynamics (cf. \citealt{Mateo1998,Strigari2008,Wolf2010,Tollerud2012,Toloba2014}) cannot be explained with CDM, since TDGs contain no CDM (cf. \citealt{Barnes1992a,Bournaud2010}).

Given the absence of CDM in dEs if they are old TDGs, we investigate how well the internal dynamics of early type galaxies (ETGs) can be understood without CDM. To this end, we test in this paper how the dynamics of ETGs would be affected if the assumption that they contain CDM is replaced by an alternative assumption, or a combination of alternative assumptions. 

The first assumption that we consider is that the ETGs obey Milgromian dynamics (MOND) instead of Newtonian dynamics. Milgromian dynamics has been suggested more than 30 years ago by \citet{Milgrom1983a} as a modification to Newtonian dynamics in the limit of very weak space-time curvature. So far, Milgromian dynamics has been very successful in quantifying the internal dynamics of galaxies without dark matter \citep{Sanders2002,Famaey2012}, even though it still has problems on larger scales (see discussion in \citealt{Famaey2012}). There is a long history of studying the detailed dynamics of elliptical galaxies in Milgromian dynamics, from the most massive ellipticals \citep{Milgrom2003,Schuberth2006,Tiret2007,Milgrom2012} to the least massive ones \citep{McGaugh2010,McGaugh2013b,Lueghausen2014}, showing globally quite good an agreement. In particular, scaling relations for the dark halos inferred around ellipticals in Newtonian dynamics have been shown to conform to Milgromian expectations \citep{Richtler2011}. However, studies of individual ETGs also reveal that some problems can occur when the dynamics of ETGs is modelled only with Milgromian dynamics. For instance, \citet{Samurovic2014} considered a sample of 10 ETGs, and found that he can explain with his models the dynamics in the centres of 9 of them without dark matter, if he assumes Milgromian dynamics instead of Newtonian dynamics. However, his models still imply dark matter at larger radii for 5 out of these 9 ETGs. As an example for an ETG that can be modelled well with Milgromian dynamics instead of CDM, we mention NGC 5128 \citep{Samurovic2010}, while NGC 1399 is an example for a galaxy where models with Milgromian dynamics still imply the presence of dark matter \citep{Richtler2008,Samurovic2016}. However, \citet{Milgrom2012} showed that in isolated ellipticals surrounded by a hot gas corona, his Milgromian predictions based on the hydrostatic equilibrium assumption are actually broadly correct in NGC 1521 out to no less than 200 kpc (corresponding to more than 10 $R_{\rm e}$ according to data on that galaxy in the NASA/IPAC Extragalactic Database\footnote{http://ned.ipac.caltech.edu/}), while problematic galaxies such as NGC 1399 tend to reside in the centre of galaxy clusters where additional missing mass is known to be needed (e.g. \citealt{Angus2008}). While many of the above studies test the gravitational potential of ETGs out to several effective radii, and also in great detail by taking, for instance, different assumptions on anisotropy into account, they also deal only with rather small samples of ETGs, or even individual ETGs. The reason is that these approaches usually rely on a sufficiently large number of objects that trace the potential of the studied galaxies, like a large population of globular clusters or a large number of planetary nebulae (or alternatively on large amounts of hot gas). The necessary data on the tracer population can only be obtained in rather expensive observations, if a suitable population exists at all around a given galaxy. For instance, the typical sizes of populations of globular clusters increase roughly with the luminosities of their host ETGs, which implies that low-mass ETGs only have a few GCs, if any. This is exemplified with the low-mass ETGs in the Local Group, which are mostly accompanied by no globular clusters, in contrast to massive ETGs like NGC 1399 or NGC 4486, which are accompanied by hundreds or even thousands of globular clusters. We therefore take here a different approach, which is to base our Milgromian estimate for each ETG on quantities that characterise every ETG and are relatively easy to observe. This allows us to consider a very large sample of ETGs that covers the whole luminosity range of ETGs, and use it to study the general trend of the Milgromian predictions, even though our estimate for each individual ETG is not very accurate. Another important caveat is that our approach is also not sensitive to the dynamics in the outskirts of the ETGs, in contrast to the more detailed studies mentioned above. 

The second assumption is that we consider an influence of binary stars on the observed internal dynamics of a galaxy. A possible effect of binaries is often neglected in estimates of the mass of galaxy from its internal dynamics. However, binaries increase the observed internal velocity dispersions of galaxies, and thus lead to too high mass estimates from the internal dynamics if binaries are neglected \citep{Kouwenhoven2008}. We note that \citet{Kouwenhoven2008} discuss star clusters instead of ETGs. However, the underlying problem is in either case that the observed motion of a single star is only due to the gravitational potential of the star cluster or the ETG to which it is bound, while the observed motion of the constituent stars of a binary has two components. The first component to the motion of these constituents is their motion as a gravitationally bound system around their common centre of mass. The second component is the motion of the centre of mass of the binary within the gravitational potential of the star cluster or the ETG. However, star clusters are in this context more complicated than ETGs, since star clusters typically have relaxation times lower than a Hubble time and can thus be dynamically evolved, while this can be excluded for ETGs \citep{Forbes2011}. This means in particular that star clusters can become mass segregated, i.e. that more massive bodies slow down and tend to gather in the centre of the evolving star cluster due to energy equipartition \citep{Binney1987}. This implies that the dynamics of more massive bodies becomes different to that of lighter bodies, and binaries as systems consisting of two stars are indeed expected to be more massive than single stars. Nevertheless, fig.~9 in \citet{Kouwenhoven2008} suggests that in stellar systems that approach masses and effective radii of ETGs, the internal dynamics of binaries becomes an important component of the observed dynamics of the system, i.e. the combination of the internal dynamics of the binaries and their dynamics as systems within the potential of their hosts. It can therefore be concluded that the results by \citet{Kouwenhoven2008} indicate that a population of binaries is also of relevance for our study on ETGs.
 
This paper is organised as follows. In Section~\ref{sec:data}, we give a brief description of the data that we use. In Section~\ref{sec:methods}, it is described how the internal velocity dispersions of ETGs are estimated from the amount of baryonic matter they contain, assuming either Newtonian dynamics or Milgromian dynamics. It is also described in that section how binaries affect the observed internal velocity dispersion of ETGs, and it is discussed how mass estimates appearing in this paper shall be interpreted. In Section~\ref{sec:results}, the methods described in Section~\ref{sec:methods} are used to quantify the expected internal dynamics and the resulting mass estimates for a sample of over 1500 ETGs, from giant elliptical galaxies to dwarf spheroidal galaxies. The calculated velocity dispersions are compared to observed values. In Section~(\ref{sec:conclusion}), we give our conclusions.

\section{Data}
\label{sec:data}

\subsection{Selection criteria for the sample of ETGs}
\label{sec:selection}

For a comparison between the theoretical predictions of the internal velocity dispersions in early-type galaxies (ETGs) with observed values, an appropriate sample of ETGs is needed. In this paper, the catalog of ETGs compiled by \citet{Dabringhausen2016} is used. This catalog comprises 1715 ETGs, which span the whole luminosity range of ETGs from faint dwarf spheroidal galaxies to giant elliptical galaxies. The quantities from that catalogue that are relevant for the present paper are the effective half-light radii of the ETGs, $R_{\rm e}$, their S\'{e}rsic indices, $n$, their observed central line-of-sight (LOS) velocity dispersions, $\sigma_{\rm obs}$, and the masses of their stellar populations, $M_{\rm s}$. We also collect their $V$-band luminosities, $L_V$, which are relevant for a discussion of the expected $V$-band mass-to-light ratios ($M/L_V$) of the galaxies. We require for all ETGs that we select from the catalogue by \citet{Dabringhausen2016} that at least $M_{\rm s}$, $R_{\rm e}$, $n$ and $L_V$ are given, while a value for $\sigma_{\rm obs}$ can be missing. This results to a sample of 1559 ETGs, and for 723 of them we also have an estimate for their $\sigma_{\rm obs}$ from the catalogue by \citet{Dabringhausen2016}. However, with some assumptions that are introduced in Section~\ref{sec:methods}, estimates on the expected LOS velocity dispersions of ETGs can be made based on their $M_{\rm s}$, $R_{\rm e}$ and $n$, and thus for every galaxy considered in this paper due to our selection criteria on ETGs from the catalogue by \citet{Dabringhausen2016}.

While we refer the reader to \citet{Dabringhausen2016} for a detailed description of their catalogue, we also give in the following some basic information on the data in their catalogue are relevant for the present paper.

\subsection{Effective half-light radii}
\label{sec:data-radii}

The effective half-light radii, $R_{\rm e}$, published in \citet{Dabringhausen2016} are based on observed values from a multitude of papers and have been homogenised for their catalogue. If angular radii are published in their sources, they transform them into pc using the distance estimates they adopted for the galaxies in their catalogue. Values for $R_{\rm e}$ given in pc in the source papers are transformed by \citet{Dabringhausen2016} into angular radii using the distance estimates given in the source papers, and then transformed back into pc again using the respective distance estimates listed in \citet{Dabringhausen2016}.

\subsection{S\'{e}rsic indices}
\label{sec:data-sersic}

The S\'{e}rsic indices, $n$, given in the catalogue by \citet{Dabringhausen2016} are adopted from the literature they used, if they are available there. If not, they calculate $n$ from $R_{\rm e}$ using
\begin{equation}
\log_{10} (n) = 0.28+0.52 \log_{10}\left(\frac{R_{\rm e}}{{\rm kpc}}\right),
\label{eq:sersic}
\end{equation}
which is a relation that \citet{Caon1993} derived from observational data.

\subsection{Velocity dispersions}
\label{sec:data-velocities}

In the context of the present paper, we have to distinguish carefully between the LOS velocity dispersion of the tracer population due to the gravitational potential of the ETG, and the observed LOS velocity dispersion. We denote the former as $\sigma_0$, and the latter as $\sigma_{\rm obs}$. The two quantities are often used interchangeably, i.e. it is assumed that $\sigma_{\rm obs}$ indicates the motion of the tracer population due to the gravitational potential of their host system. However, this is conceptually wrong if the tracer population are binaries, i.e. systems with internal dynamics. Nevertheless, setting $\sigma_{\rm obs}=\sigma_0$ is an extremely good approximation for ETGs above a certain luminosity limit, as will be discussed in Section~\ref{sec:velocities}.  

\citet{Dabringhausen2016} follow the usual convention by designating the central LOS velocity dispersions of the ETGs as $\sigma_0$, while the quantities that they actually list in their catalogue are values for $\sigma_{\rm obs}$ according to the above definitions for $\sigma_0$ and $\sigma_{\rm obs}$. Their data on $\sigma_{\rm obs}$ are taken from the literature they use, if such a value was published there. If only the average observed LOS velocity dispersion within $R_{\rm e}$, $\sigma_{\rm e}$, is available from the literature they use, they estimate $\sigma_{\rm obs}$ using the relation
\begin{equation}
\log_{10}\left(\frac{\sigma_{\rm obs}}{{\rm km/s}}\right)=1.0478 \, \log_{10}\left(\frac{\sigma_{\rm e}}{{\rm km/s}}\right)-0.0909,
\label{eq:sigmae}
\end{equation}
which is obtained from a least-squares fit to data on 260 ETGs in \citet{Cappellari2013}, while taking the data on the velocity dispersions within the central parsec of the ETGs in \citet{Cappellari2013} as a measure for $\sigma_{\rm obs}$. By construction, equation~({\ref{eq:sigmae}}) cannot be used on ETGs with $R_{\rm e}<1$ kpc, which is why \citet{Dabringhausen2016} set $\sigma_{\rm obs}=\sigma_{\rm e}$ for such ETGs. This is well motivated by the findings that $R_{\rm e}=1$ kpc corresponds to $n \approx 1.9$ according to equation~(\ref{eq:sersic}), and that the velocity dispersion profiles of galaxies with $n\apprle 2$ are almost flat within their $R_{\rm e}$ \citep{Graham1997,Simonneau2004}.
 
\subsection{Masses of the stellar populations} 
\label{sec:data-masses}
 
The estimates on $M_{\rm s}$ given in \citet{Dabringhausen2016} are based on a large set of models for simple stellar populations (SSPs) by \citet{Bruzual2003}. These SSPs are defined as stellar populations that have formed instantly at a certain time with a certain metallicity. More specifically, \citet{Dabringhausen2016} obtain $M_{\rm s}$ of a given ETG by first searching an SSP-model by \citet{Bruzual2003} that represents the age and the colours of the ETG the best, then adopting the $M/L_V$ predicted by the according SSP-model as the $M/L_V$ of the ETG, and finally multiplying this $M/L_V$ by the $L_V$ of the ETG. The colours of the ETGs serve in this context as indicators for the metallicities of their stellar populations, since the metallicity determines the colours of an SSP with a given age, while colours are much easier to observe than metallicities.

The data on the ages of the stellar populations of the ETGs that \citet{Dabringhausen2016} use in the estimates of $M_{\rm s}$ come from the literature that they consider, or are the average value of galaxies with similar $L_V$, if no age was published for a given ETG. Also the data on the colours that \citet{Dabringhausen2016} require for their estimates of $M_{\rm s}$ comes from the literature that they use, if such data is published there. If no data was published there, \citet{Dabringhausen2016} adopt colours that are typical for ETGs with similar $L_V$. There are also cases where only $L_V$ is available for an ETG that \citet{Dabringhausen2016} consider. In that case, both age and colours are set to values that are typical for ETGs of that $L_V$.

Note that the ages considered for the stellar populations of the ETGs are characteristic ages, since the stellar populations of ETGs do not form instantly (see for instance \citealt{Thomas2005} and \citealt{Weisz2014}). As a consequence, they are composed of stars with different metallicities, so that also the metallicities that are adopted for the stellar populations of ETGs are effective values that are supposed to characterise them with a single number. This implies that in reality the dependency between metallicities and colours of real stellar populations in ETGs is not necessarily as clear as suggested by SSP-models.

Another important aspect about the values for $M_{\rm s}$ taken from the catalogue by \citet{Dabringhausen2016} is that in the SSP-models on which they are based, the stellar initial mass function (IMF) is assumed to be canonical. According to the canonical IMF, the number of newly forming stars as a function of their mass is a power law with the Salpeter slope ($\alpha=-2.3$) for stellar masses above $0.5 \ {\rm M}_{\odot}$, but with a flatter slope ($\alpha=-1.3$) for lower stellar masses. This canonical IMF is consistent with many observations in the Solar neighbourhood, but there is also evidence for deviations from the canonical IMF in more distant stellar populations (see \citealt{Kroupa2013} for a review on the IMF).

\subsection{$V$-band luminosities}
\label{sec:data-luminosities}

The data on $L_V$ in the catalogue by \citet{Dabringhausen2016} are either based on direct observations of the individual ETGs, or are derived from these data using their statistical properties. \citet{Dabringhausen2016} prefer data for $L_V$ based on direct observations, which they obtain from apparent $V$-band magnitudes collected from many sources in the literature and the distance estimates they adopt for the according galaxies. They perform some basic homogenisation for this type of data, using overlaps between the different samples of galaxies introduced in the literature that they use. This is done in order to take care of offsets between the different data samples, which most likely originate from observations with different telescopes at different times, and different procedures for data reduction. If no published value for the apparent $V$-band magnitude of a galaxy was available from the literature, while there was such data from neighbouring passbands, \citet{Dabringhausen2016} use linear relations between $L_V$ and luminosities in those other passbands in order to estimate $L_V$ of the galaxy.

It turns out that the availability of $L_V$ is the bottleneck for our selection of ETGs from the catalogue by \citet{Dabringhausen2016} due to its key role for estimating $M_{\rm s}$, so that in the end only 1559 ETGs of the 1715 ETGs listed in the catalogue by \citet{Dabringhausen2016} are considered here. \citet{Dabringhausen2016} found data on $R_{\rm e}$ for all ETGs in their catalogue but two, and thus they also give values for $n$ for them, which are calculated by equation~(\ref{eq:sersic}) if they found no previously published value for $n$. However, an estimate for $M_{\rm s}$ by the methods in \citet{Dabringhausen2016} is impossible without knowing $L_V$, while \citet{Dabringhausen2016} give an estimate for $M_{\rm s}$ also if only $L_V$ is known, even though these estimates are by construction inferior to the estimates where they have additional observed parameters at their disposal.

\section[Methods]{Methods}
\label{sec:methods}

\subsection{Predictions for the velocity dispersions in ETGs}
\label{sec:MOND}

\subsubsection{General assumptions}
\label{sec:assumptions}

With some additional assumptions apart from the ones \citet{Dabringhausen2016} made for their catalogue (see Section~\ref{sec:data}), the central velocity dispersion of an ETG is here estimated from its $M_{\rm s}$, $R_{\rm e}$ and $n$.

The first assumption is that the ETGs are not embedded in haloes of non-baryonic dark matter and that moreover mass follows light in them. 

The second assumption is that $M_{\rm s}$, i.e the mass of the matter that is locked up in stars and stellar remnants, is indeed a good approximation for $M$, which denotes the mass of the total baryonic matter contained in the ETG. This assumption is motivated by the findings that ETGs contain little gas \citep{Young2011} and dust \citep{Dariush2016}. Defining $M_{\rm N}$ as the mass that is projected within $R_{\rm e}$, the combination of this assumption with the previous assumption implies that 
\begin{equation}
M_{\rm N}=\frac{M}{2}=\frac{M_{\rm s}}{2},
\label{eq:MNEWT}
\end{equation}
i.e. that $M_{\rm N}$ is half of the matter contained by the ETG, since the first assumption then implies that the stellar population does not vary throughout the ETG. Whether $M_{\rm N}$ indeed accounts well for the baryonic matter in the ETG is however also a question of how well the assumed SSP approximates the properties of the actual stellar population of the ETG. With a poor choice for the SSP, the estimate of $M_{\rm N}$ may be wrong by up to a factor of a few. The importance of the choice of the SSP (most notably the IMF with which it formed) is for instance illustrated in table~5 in \citet{Samurovic2014}. However, even for a fixed IMF, different ages and metallicities can imply substantial changes in $M_{\rm N}$ for a given luminosity. This can be seen for instance in the data by \citet{Bruzual2003}, who provide the SSP-models on which our estimates of $M_{\rm s}$ and thus $M_{\rm N}$ are based. The combination of the first and the second assumption also implies that the composition of the stellar population does not change significantly spatially within the ETG.

The third assumption is that the density profiles of the ETGs can be approximated as spherically symmetric. As a consequence of this assumption, we use $R_{\rm e}$ in the following for estimating the masses and LOS velocity dispersions of the ETGs.

The fourth assumption is that the anisotropy parameter, $\beta$, of the galaxy is close to zero (cf. equation~4-53b in \citealt{Binney1987}). This means that systematic patterns in the motions of the stars of the galaxy can be neglected. As a consequence, the mass of the ETG will be overestimated if $-\infty<\beta<0$ (i.e. if the stellar orbits in the ETG are predominately circular) and underestimated if $0<\beta<1$ (i.e. if the orbits in the ETG are predominately radial).

The fifth assumption is that the ETG is in virial equilibrium. This means that the ETG has settled into a configuration where its gravitational potential and thus its density profile does not depend on time. This assumption implies in particular that the ETG is not significantly disturbed by tidal forces, i.e. external, time-dependent gravitational fields. Tidal fields increase the actually observed internal velocity dispersions and may even disrupt galaxies \citep{Kroupa1997,Fellhauer2006,McGaugh2010,Casas2012,Yang2014,Dominguez2016}.

Finally, as a sixth assumption, a law of gravity that governs the internal dynamics of the ETG has to be adopted. We discuss the case of Newtonian dynamics in Section~\ref{sec:newton} and the case of Milgromian dynamics in Section~\ref{sec:milgrom}.

\subsubsection{Newtonian dynamics}
\label{sec:newton}

If Newtonian dynamics is adopted, and the other assumptions given in Section~\ref{sec:assumptions} are justifiable as well, the total mass of the ETG can be approximated with
\begin{equation}
M=M_{\rm s}=\frac{K_{\rm V}}{G}R_{\rm e} \sigma_{0}^2,
\label{eq:Mdyn}
\end{equation}
which is a relation derived by \citep{Bertin2002} using the virial theorem (see their equation~7). In this equation, $G$ is the gravitational constant, $R_{\rm e}$ is the effective half-light radius, $\sigma_0$ is the central LOS velocity dispersion of the tracer population, and $K_{\rm V}$ is the virial coefficient. Recall that even for ETGs in virial equilibrium, $\sigma_0$ is not necessarily equal to the observed LOS velocity dispersion, $\sigma_{\rm obs}$ (see Section~\ref{sec:data-velocities}). The equality of the total mass, $M$, and the total mass of the stellar population, $M_{\rm s}$, follows from equation~(\ref{eq:MNEWT}), and thus from the assumptions that we make for our study. The value for $K_{\rm V}$ depends on the S\'{e}rsic index $n$ as
\begin{equation}
K_{\rm V}(n) = \frac{73.32}{10.465+(n-0.94)^2}+0.954
\label{eq:virialcoeff}
\end{equation}
according to equation~11 in \citet{Bertin2002}.

In equations~(\ref{eq:Mdyn}) and~(\ref{eq:virialcoeff}), $M$ (which is  equal to $M_{\rm s}$ in the context of the present paper) is treated as an unknown parameter that can be estimated from observed values for $R_{\rm e}$, $n$ and $\sigma_0$. We now consider the case that $M_{\rm s}$, $R_{\rm e}$ and $n$ are known parameters taken from the catalogue by \citet{Dabringhausen2016}, while $\sigma_{\rm N}$, i.e. the LOS velocity dispersion based on the assumption of Newtonian dynamics, is treated as the unknown parameter. In order to estimate $\sigma_{\rm N}$, we replace $\sigma_0$ in equation~(\ref{eq:Mdyn}) with $\sigma_{\rm N}$, and solve the equation for the unknown parameter. This yields
\begin{equation}
\sigma_{\rm N}=\sqrt{\frac{2\, G \, M_{\rm N}}{K_{\rm V} \, R_{\rm e}}},
\label{eq:sigmaNEWT}
\end{equation}
where also equation~(\ref{eq:MNEWT}) was used in order to replace $M$ with $M_{\rm N}$.

\subsubsection{Milgromian dynamics}
\label{sec:milgrom}
 
If Milgromian dynamics instead of Newtonian Dynamics is assumed, another estimator is needed. One common choice for faint systems is based on the ``deep-MOND virial relation" \citep[e.g.][]{McGaugh2013b}. However, given the wide range of internal accelerations probed by our sample, this estimator is not valid in most systems, and we thus take here a less precise but more general approach based on the ``phantom dark matter" distribution (see below).

The general idea of Milgromian dynamics is that, at the most basic level in highly symmetric systems, the gravitational acceleration, $\mathbf{a}$ is related to the gravitational acceleration according to Newtonian Dynamics, $\mathbf{a}_{\rm N}$ by
\begin{equation}
\mathbf{a}_{\rm N}=\mu(a/a_0) \ \mathbf{a},
\label{eq:acc}
\end{equation}
where $a_{0}$ is the critical acceleration \citep{Milgrom1983a}. Below $a_{0}$, the gravitational force acting on a test particle is for a given matter distribution stronger than predicted by Newtonian dynamics. Note that the gravitational accelerations within the Solar system are higher than $a_{0}$ by four orders of magnitudes (cf. fig.~11 in \citealt{Famaey2012}). 

The transition between the regime of Newtonian dynamics and Milgromian dynamics is given by an interpolating function that has to satisfy
\begin{equation}
\mu(a/a_0) \rightarrow 1 \quad {\rm for} \quad a\gg a_{0}
\end{equation}
and 
\begin{equation}
\mu(a/a_0) \rightarrow a \quad {\rm for} \quad a\ll a_{0}.
\end{equation}
Even though gravitational accelerations in the Solar System are high compared to $a_0$, this interpolating function is tightly constrained in its strong field limit from very precise constraints on the precession of Saturn perihelion obtained by Cassini radioscience tracking data \citep{Hees2016}. However, the shape of the interpolating function in the range of accelerations probed by galaxies is less constrained, and according to \citet{Famaey2005} and \citet{Zhao2006}, a possible choice for the interpolating function there is the simple $\mu$-function,
\begin{equation}
\label{eq:mu}
\mu(a/a_0)=\frac{(a/a_0)}{1+(a/a_0)}.
\end{equation}
Inserting equation~(\ref{eq:mu}) into equation~(\ref{eq:acc}) and solving for $a$ leads to
\begin{equation}
\label{eq:MassMOND}
a=\frac{a_{\rm N}}{2}\left(1+\sqrt{1+\frac{4\, a_0}{a_{\rm N}}}\right),
\end{equation}
which relates the acceleration according to Milgromian Dynamics to the acceleration according to Newtonian Dynamics. Following \citet{Famaey2005}, we set $a_{0}=1.2\times 10^{-10} \rm{m \, s^{-2}}$ in our paper.

More generally, Milgromian dynamics can be formulated such that the generalised Poisson equation is given as
\begin{equation}
\nabla^2\Phi=4\pi G \left[\rho_{N}+\frac{1}{8 \pi G}\nabla \cdot \left[ \left( \sqrt{1+\frac{4\, a_{0}} {a_{\rm N}}}-1 \right) {\mathbf a_{\rm N}} \right] \right],
\end{equation}
where the first term within the brackets on the right is the density of actual matter, $\rho_{\rm N}$, and the second term follows from the potential this matter would create according to Newtonian Dynamics \citep{Milgrom2010,Lueghausen2014}. While this second term does not correspond to the density of a type of actual matter, it can be interpreted as the density of ``phantom dark matter", $\rho_{\rm PDM}$ \citep{Lueghausen2014}. Taking the sum of $\rho_{\rm N}$ and $\rho_{\rm PDM}$ as the apparent matter density in Milgromian Dynamics, $\rho_{\rm M}$, the Poisson equation assumes in Milgromian Dynamics the same form that is known from Newtonian Dynamics, but with $\rho_{\rm M}$ instead of $\rho_{\rm N}$. On this basis, problems in Milgromian dynamics can be treated mathematically like problems in Newtonian dynamics, once $\rho_{\rm M}$ is known. Hence, $M_{\rm M}$ can in principle be found by integrating $\rho_{\rm M}$ over a volume and the velocity dispersion of a system that obeys Milgromian dynamics can be estimated with the equations that are known from Newtonian dynamics, if $M_{\rm N}$ is replaced with $M_{\rm M}$.

In pure spherical symmetry, the Newtonian gravitational acceleration at the radius $R$ of a spherical mass distribution is given as
\begin{equation}
a_{\rm N}(R)=\frac{G \, M(R)}{R^2},
\label{eq:NewtAcc1}
\end{equation}
where $M(r)$ is the total mass within the radius $R$ and $G$ is the Newtonian gravitational constant. At the radius $R_{\rm e}$, $a_{\rm N}$ can be approximated with 
\begin{equation}
a_{\rm N}=\frac{G \, M_{\rm N}}{R_{\rm e}^2}.
\label{eq:NewtAcc2}
\end{equation}
Inserting equation~(\ref{eq:NewtAcc2}) into equation~(\ref{eq:MassMOND}) leads to
\begin{equation}
a=\frac{G}{2} \frac{M_{\rm N}}{R_{\rm e}^2}\left(1+\sqrt{1+\frac{4\, a_0}{G} \frac{R_{\rm e}^2}{M_{\rm N}}}\right). 
\label{eq:MassMOND1}
\end{equation}
Equation~(\ref{eq:NewtAcc2}) also implies that in Newtonian Dynamics the mass $M_{\rm N}$ enclosed within the radius $R_{\rm e}$ can be expressed in terms of $R_{\rm e}$ and the acceleration $a_{\rm N}$ acting on a test particle at $R_{\rm e}$. In analogy to this, an apparent mass according to Milgromian dynamics can be introduced as
\begin{equation}
M_{\rm M}= \frac{a \, R_{\rm e}^2}{G}=\frac{M_{\rm N}}{2}\left(1+\sqrt{1+\frac{4\, a_0}{G} \frac{R_{\rm e}^2}{M_{\rm N}}}\right),
\label{eq:MassMOND2}
\end{equation}
where equation~(\ref{eq:MassMOND1}) was used for the last equality. Based on this estimate for the apparent mass, the prediction for the central LOS velocity dispersion in a ETG in Milgromian dynamics is
\begin{equation}
\sigma_{\rm M}=\sqrt{\frac{2\, G \, M_{\rm M}}{K_{\rm V} \, R_{\rm e}}},
\label{eq:sigmaMOND}
\end{equation}
which is analogous to equation~(\ref{eq:sigmaNEWT}) for Newtonian dynamics. We also have to assume that most of our systems will not be affected by the external field effect (EFE) of Milgromian dynamics. Although this could be an issue for some of the faintest systems \citep{McGaugh2013b} and those close to the centre of galaxy clusters, it is safe to assume that this will not affect the majority of our systems and the general trends \citep[e.g.][]{Richtler2011}.

\subsection{The effect of binaries on observed velocity dispersions}
\label{sec:binaries}

The catalogue by \citet{Dabringhausen2016} contains observed central LOS velocity dispersions, $\sigma_{\rm obs}$, for many of the galaxies for which we predict $\sigma_{\rm N}$ and $\sigma_{\rm M}$ here. For a comparison between $\sigma_{\rm obs}$ and $\sigma_{\rm N}$ or $\sigma_{\rm M}$, it has to be considered that $\sigma_{\rm N}$ and $\sigma_{\rm M}$ only quantify the velocity dispersion of a tracer population within the potential of the ETG. Suspecting similar conditions like among field stars in the Milky Way, the tracer population that is used to infer $\sigma_{\rm obs}$ is not a population of only single stars, but a mixture of single stars and multiple systems, of which most are binaries \citep{Abt1976,Duquennoy1991}. Thus, some members of the tracer population in actual ETGs possess an internal structure that influences $\sigma_{\rm obs}$. We seek to quantify this influence in the following.

The orbital motions of stars in a binary system are uncorrelated with the orbital motions of the binary systems (and single stars) in the potential of the ETG. Thus, assuming that all stars are either single stars or part of a binary, the influence of binary stars on $\sigma_{\rm obs}$ can be quantified as
\begin{equation}
\sigma_{\rm obs}^2=(1-f)\sigma_0^2+f(\sigma_0^2+\sigma_{\rm bin}^2)=\sigma_0^2+f\sigma_{\rm bin}^2,
\label{eq:binaries}
\end{equation}
where $f$ is the fraction of binaries among all centre-of-mass systems (i.e. single stars and binaries), $\sigma_0$ is the velocity dispersion of the centre-of-mass systems due to virialised motion in the potential of the ETG, and $\sigma_{\rm bin}$ quantifies the LOS component of the typical orbital velocity of the stars in binaries.

\citet{Gieles2010} derive an analytic approximation for $\sigma_{\rm bin}$, which can be formulated as
\begin{equation}
\sigma_{\rm bin}^2=\frac{1}{3}\left(\frac{2q^{3/2}}{1+q}\right)^{4/3} \left(\frac{\pi G m_1}{2P}\right)^{2/3},
\label{eq:sigma-bin}
\end{equation}
where $q$ is the mass ratio of the two components of the binary, $m_1$ is the typical mass of the more massive component of the binary and $P$ is the typical orbital period in the binaries.

In order to find an appropriate value for $\sigma_{\rm bin}$, and consequently $\sigma_{\rm obs}$, the properties of the binary populations in ETGs have to be constrained. As the ETGs have old stellar populations, the binaries that are relevant here are those, where the components are stars with masses of about one Solar mass or less, i.e. G-stars or M-stars. \citet{Duquennoy1991} found that the distribution of orbital periods of G-dwarf binaries peak at $P \approx$ 180 yr, and that the binary fraction for these systems is $f \approx 0.4$ when no constraints on $P$ are made. \citet{Minor2013} finds that the Fornax, Sculptor and Sextans dSphs are consistent with having similar populations of binaries as the binary population that \citet{Duquennoy1991} observed in the Galactic field, while the Carina dSph either contains fewer binaries or binaries with wider separations. Relevant for an increase of $\sigma_{\rm obs}$ due to binaries are in particular the close binaries, since their components have higher orbital velocities \citep{McConnachie2010}. For rather tight binaries (i.e. binaries with $P \apprle 20$ yr), observations suggest $0.2 \apprle f \apprle 0.3$  \citep{Carney2005,Marks2011}. In this paper, we use in equation~(\ref{eq:sigma-bin}) $q=0.6$ (cf. \citealt{Gieles2010}), $P=10$ yr and $m_1 = 1 \ {\rm M}_{\odot}$, which leads to $\sigma_{\rm bin} = 2.3$ km/s. In equation~(\ref{eq:binaries}), we consider the cases $f=0$ (i.e. no binaries) and $f=0.3$.

In the following, predictions for $\sigma_{\rm obs}$ made with equation~(\ref{eq:binaries}) for $f=0.3$ will be denoted as $\sigma_{\rm N}^{\rm bin}$ if $\sigma_0$ in equation~(\ref{eq:binaries}) is set to $\sigma_{\rm N}$, and as $\sigma_{\rm M}^{\rm bin}$, if $\sigma_0$ in equation~(\ref{eq:binaries}) is set to $\sigma_{\rm M}$. For $f=0$, binaries have no impact on the velocity dispersions, and they are denoted as $\sigma_{\rm N}$ for the Newtonian case and  as $\sigma_{\rm M}$ for the Milgromian case.  As a summarising term for different predictions for $\sigma_{\rm obs}$, we will use $\sigma_{\rm pred}$.

The number we calculate with equation~(\ref{eq:sigma-bin}), and consequently the estimates on $\sigma_{\rm obs}$ with equation~(\ref{eq:binaries}) are no more than rough estimates. However, given the current knowledge on binary populations in ETGs \citep{McConnachie2010,Minor2013}, it seems unlikely that the effect of binaries on $\sigma_{\rm obs}$ will be much larger than what we consider with our assumptions. In any case, we note that the exponents in equation~(\ref{eq:sigma-bin}) rather close to unity, so that the estimate of $\sigma_{\rm bin}$, and the predictions for $\sigma_{\rm obs}$ of ETGs are of the right order of magnitude, if the values for the parameters $f$, $q$, $m_1$ and $P$ are.

We also note that in equation~(\ref{eq:sigma-bin}), binaries are considered as systems that obey Newtonian dynamics. It may therefore seem inconsistent to use $\sigma_{\rm bin}$ to estimate $\sigma_{\rm obs}$ with equation~(\ref{eq:binaries}) in a system, where the motion of the centre-of-mass systems is assumed to obey to Milgromian dynamics. However, we assume $P=10$ yr for the typical orbital period in binaries, and if the binary system has a total mass of $1 \, {\rm M}_{\odot}$, $P=10$ yr corresponds to a semi-major axis of almost 5 AU with Kepler's 3rd law. This is similar to the semi-major axis of the orbit of Jupiter, where the gravitational acceleration due to the Sun is still larger than $a_{\rm 0}$ by about than 5 orders of magnitude, as illustrated in fig.~11 in \citet{Famaey2012}. 

Another important aspect about equation~(\ref{eq:sigma-bin}) is that it neglects the contribution of the less luminous component of the binary to $\sigma_{\rm bin}$. \citet{Gieles2010} justify this by noting that the more massive component of the binary is also the more luminous, and thus has the most impact on $\sigma_{\rm obs}$. This is a safe assumption also for the case of binaries in ETGs. The most important tracers for $\sigma_{\rm obs}$ in an ETG are stars with masses close to $1 {\rm M}_{\odot}$, since these are the brightest stars that have not evolved into stellar remnants yet in old stellar populations, as they are usually found in ETGs. If such stars are one of the components in a binary, they are also very likely to be the more massive object in it, since a likely companion for a star with a mass close to $1 {\rm M}_{\odot}$ in an ETG is either a less massive star that is still on the main sequence, or a white dwarf. Both these types of companions are certainly the less luminous component of the binary, and also white dwarfs are usually less massive than $1 {\rm M}_{\odot}$, even if they originate from stars that were initially more massive than $1 {\rm M}_{\odot}$ (see \citealt{Kalirai2008}).

\subsection{Estimates for the mass-to-light ratios of ETGs}
\label{sec:masses}

At least formally, mass estimates for ETGs can be made with equation~(\ref{eq:Mdyn}) when $\sigma_{0}$ in equation~(\ref{eq:Mdyn}) is replaced with $\sigma_{\rm N}$, $\sigma_{\rm N}^{\rm bin}$,  $\sigma_{\rm M}$, $\sigma_{\rm M}^{\rm bin}$ and $\sigma_{\rm obs}$. In the following, the resulting quantities will be denoted as $M_{\rm N}$, $M_{\rm N}^{\rm bin}$,  $M_{\rm M}$, $M_{\rm M}^{\rm bin}$, and $M_{\rm obs}$, respectively. As a summarising term for $M_{\rm N}$, $M_{\rm N}^{\rm bin}$,  $M_{\rm M}$ and $M_{\rm M}^{\rm bin}$, we use $M_{\rm pred}$. When mass estimates are made from observational data, $\sigma_{\rm obs}$ is quite often identified with $\sigma_0$ in the literature. However, this is not strictly correct if the ETGs contain binaries, while it is generally an excellent approximation for ETGs with luminosities ${\rm L}_{\odot} \apprle 10^8 {\rm L}_{\odot}$ (see Section~\ref{sec:MLratios}).

While $M_{\rm N}$ would actually be an estimate for the real mass of an ETG, if it contains no binaries and its internal dynamics is Newtonian, the quantities $M_{\rm N}^{\rm bin}$, $M_{\rm M}$ and $M_{\rm M}^{\rm bin}$ are not. Instead, they are best understood as quantities that allow to decide how well $M_{\rm obs}$ can be understood from only the likely amount of baryons in the ETGs, given the additional assumptions laid out in Sections~\ref{sec:milgrom} and~\ref{sec:binaries}. $M_{\rm pred} \approx M_{\rm obs}$ indicates that the dynamics of the ETG can be explained without additional assumption, while a possible interpretation for $M_{\rm pred}<M_{\rm obs}$ is that the ETG contains additional, dark, but not necessarily non-baryonic matter.

For a discussion of the extreme mass-to-light ratios reported for many low-mass ETGs in the literature, we divide $M_{\rm pred}$ and $M_{\rm obs}$ by the $V$-band luminosities listed for these galaxies in the catalogue by \citet{Dabringhausen2016}.

\section[Results]{Results}
\label{sec:results}

\subsection[Velocity dispersions]{Velocity dispersions}
\label{sec:velocities}

\begin{figure*}
\centering
\includegraphics[scale=0.8]{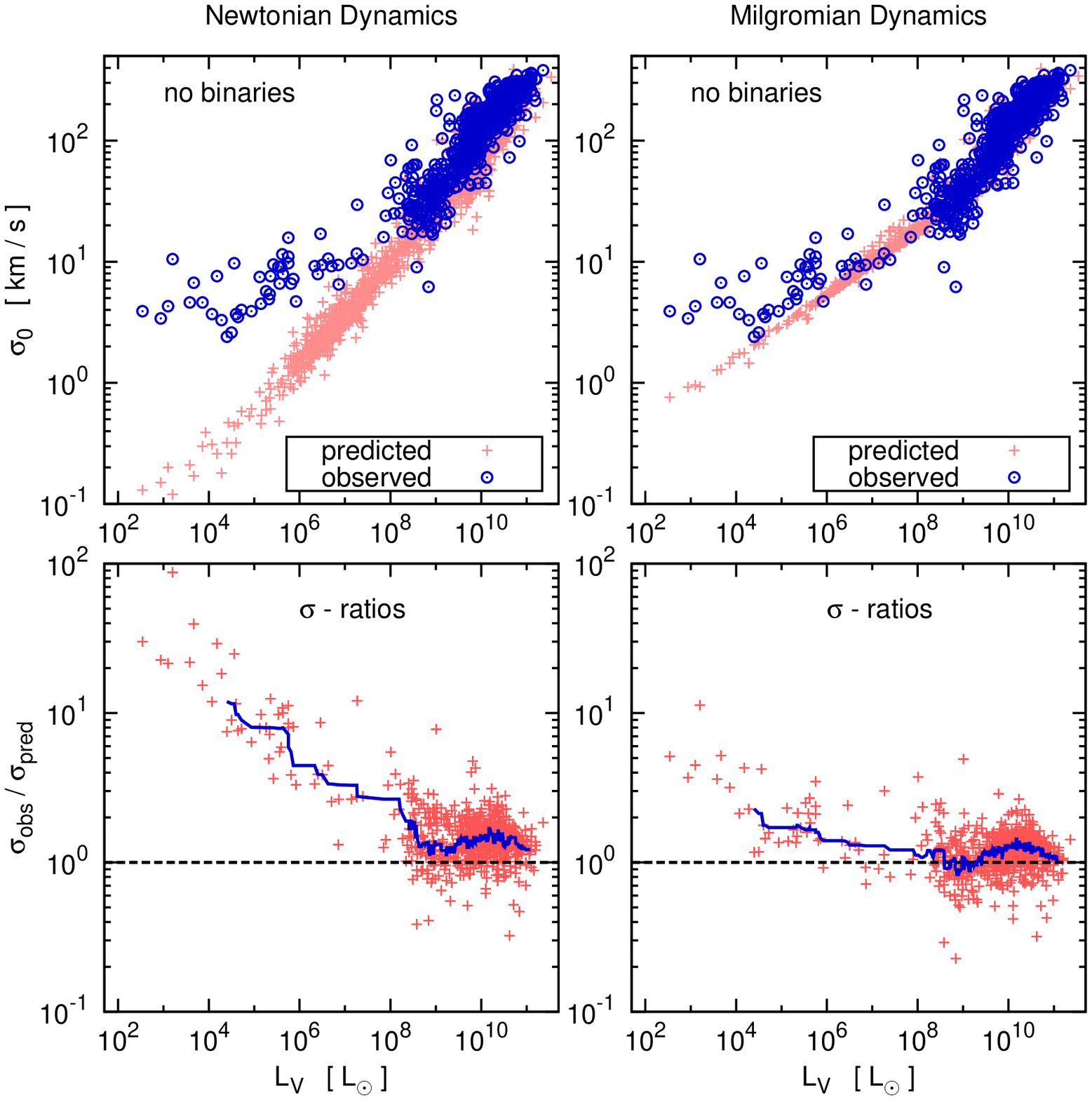}
\caption[Comparison between $\sigma_{\rm pred}$ and $\sigma_{\rm obs}$ of the ETGs, without considering a possible effect of binaries on the observed velocity dispersions.]{\label{fig:velo-no-bin} The Comparison between $\sigma_{\rm pred}$ and $\sigma_{\rm obs}$ of the ETGs, without considering a possible effect of binaries. In the panels on the left, Newtonian dynamics is assumed for calculating $\sigma_{\rm pred}$, and Milgromian dynamics in the panels on the right. In the two top panels, the values for $\sigma_{\rm obs}$ are plotted on top of the values for $\sigma_{\rm pred}$. The two bottom panels show the ratio between $\sigma_{\rm obs}$ and $\sigma_{\rm pred}$ for the ETGs where both quantities are known. The solid lines in the bottom panels quantifies change of the median value of $\sigma_{\rm obs}/\sigma_{\rm prec}$ with luminosity. Note that the method for finding these median values does not allow us to assign median values to the luminosities of the ten faintest and the ten brightest galaxies shown in the figure. The dashed lines in these panels indicate equality between the predictions and the observations.}
\end{figure*}

\begin{figure*}
\centering
\includegraphics[scale=0.8]{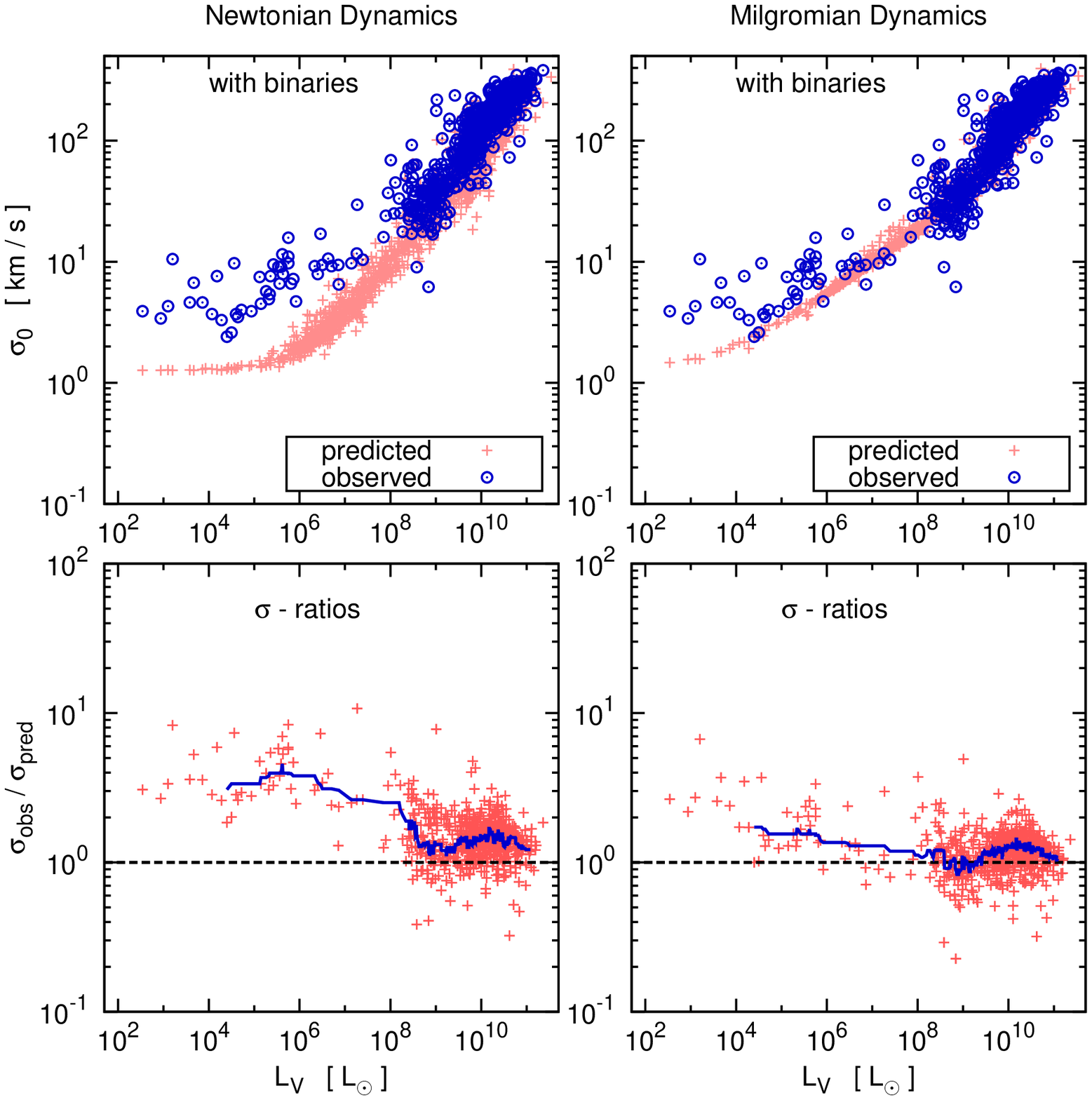}
\caption[Comparison between the predicted velocity dispersions and $\sigma_{\rm obs}$ of the ETGs, considering a possible effect of binaries on the observed velocity dispersions.]{\label{fig:velo-with-bin} As Figure \ref{fig:velo-with-bin}, but assuming that 30 per cent of the stars in the ETGs are part of a binary.}
\end{figure*}

\begin{figure*}
\includegraphics[scale=0.8]{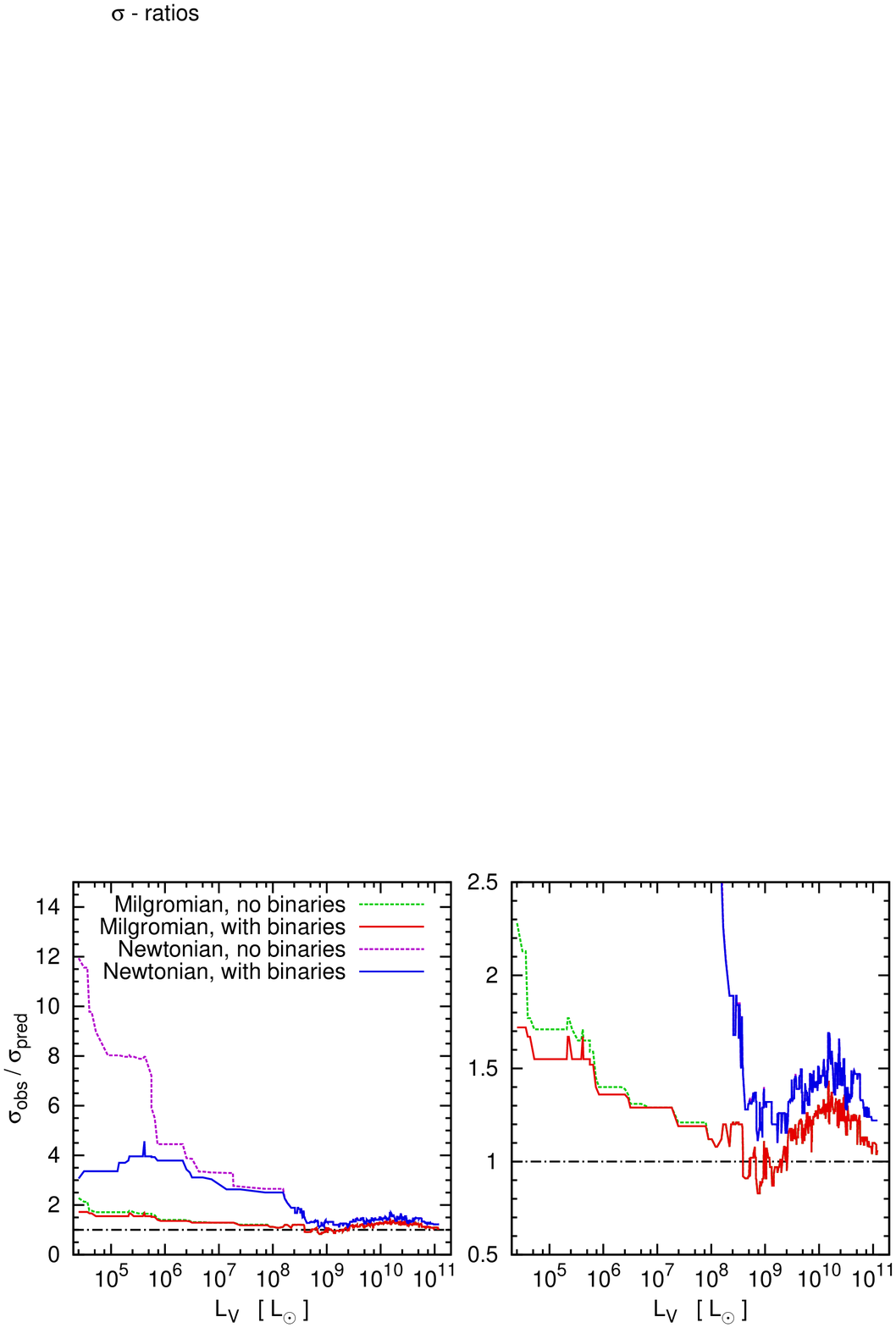}
\centering
\caption[Comparison between the medians of the ratios between $\sigma_{\rm obs}$ and the predicted velocity dispersions of ETGs.]{\label{fig:velo-comp} Comparison between the different relations quantifying the typical $\sigma_{\rm obs}/\sigma_{\rm pred}$ of ETGs, which are shown in the bottom panels of Figs.~\ref{fig:velo-no-bin} and~(\ref{fig:velo-with-bin}. The different assumptions that enter the values for $\sigma_{\rm pred}$ are specified in the left panel. The left panel shows the full range of the relations, while the right panel zooms in on the range where $\sigma_{\rm obs}$ and $\sigma_{\rm pred}$ are similar.}
\end{figure*}

Fig.~\ref{fig:velo-no-bin} shows comparisons between $\sigma_{\rm N}$ and $\sigma_{\rm M}$ with $\sigma_{\rm obs}$. Thus, a possible influence of binaries on the observed velocity dispersions of ETGs is neglected in Fig.~\ref{fig:velo-no-bin}. In the left panels, the case of Newtonian dynamics is considered (i.e. the estimated velocity dispersion is $\sigma_{\rm N}$), and in the right panels, the case of Milgromian dynamics is considered (i.e. the estimated velocity dispersion is $\sigma_{\rm M}$). 

In the two top panels of Fig.~\ref{fig:velo-no-bin}, the observed internal velocity dispersions of the ETGs, $\sigma_{\rm obs}$, are plotted on top of $\sigma_{\rm N}$ (left panel) and $\sigma_{\rm M}$ (right panel). In particular for ETGs with $V$-band luminosities $10^5 \, {\rm L}_{\odot} \apprle L_V \apprle 10^9 \, {\rm L}_{\odot}$, $\sigma_{\rm N}$ and $\sigma_{\rm M}$ can be calculated for many ETGs, for which $\sigma_{\rm obs}$ is not yet known, and is quite challenging to observe. Note that due to our selection criteria on the ETGs in the catalogue by \citet{Dabringhausen2016}, $R_{\rm e}$, $K_{\rm V}$ (which is calculated from $n$ with equation~\ref{eq:virialcoeff}) and $M_{\rm s}$ are available for all ETGs considered in the present paper, as is detailed in Section~\ref{sec:selection}. The values for $M_{\rm N}$ and $M_{\rm M}$ follow from the $M_{\rm s}$ of the galaxies with the assumptions made in Section~\ref{sec:methods}. Thus $\sigma_{\rm N}$ and $\sigma_{\rm M}$ can be calculated for all ETGs considered here with equations~(\ref{eq:sigmaNEWT}) and~(\ref{eq:sigmaMOND}), respectively.

For the ETGs for which also $\sigma_{\rm obs}$ is known, $\sigma_{\rm obs}/\sigma_{\rm N}$ is shown in the lower left panel and $\sigma_{\rm obs}/\sigma_{\rm M}$ is shown in the lower right panel of Fig.~\ref{fig:velo-no-bin}.

The solid (blue) line in the bottom left panel of Fig.~\ref{fig:velo-no-bin} serves as a quantification of a characteristic value for $\sigma_{\rm obs}/\sigma_{\rm N}$ in dependency of $L_{V}$. For obtaining this quantification, we consider all ETGs for which $\sigma_{\rm N}$ and $\sigma_{\rm obs}$ are known, and number them consecutively by ascending luminosity. From this list, we construct all possible subsets that contain 21 ETGs with consecutive numbers, i.e. all subsets that comprise all galaxies from the $i-10$th brightest to the $i+10$th brightest ETG on the list. These subsets are then sorted by $\sigma_{\rm obs}/\sigma_{\rm N}$, and the 11th highest value for $\sigma_{\rm obs}/\sigma_{\rm N}$ among this subset is then chosen for the characteristic $\sigma_{\rm obs}/\sigma_{\rm N}$ for ETGs with the luminosity of the $i$th ETG. By construction, this method cannot assign characteristic $\sigma_{\rm obs}/\sigma_{\rm N}$ to the 10 least least luminous and the 10 most luminous ETGs, but to all other ETGs. However, the advantage of this method is that the impact of ETGs with exceptional $\sigma_{\rm obs}/\sigma_{\rm N}$ on the estimates of the typical $\sigma_{\rm obs}/\sigma_{\rm N}$ is minimised.

The solid (blue) line in the right bottom panel of Fig.~\ref{fig:velo-no-bin} shows the same quantity as the solid (blue) line in the bottom left panel of that Figure, but for Milgromian dynamics instead of Newtonian dynamics.

Fig.~\ref{fig:velo-with-bin} is organised as Fig.~\ref{fig:velo-no-bin}, but in Fig.~\ref{fig:velo-with-bin}, $\sigma_{\rm N}$ is replaced by $\sigma_{\rm N}^{\rm bin}$, and $\sigma_{\rm M}$ is replaced by $\sigma_{\rm M}^{\rm bin}$. Thus, the difference between Figs.~\ref{fig:velo-no-bin} and~\ref{fig:velo-with-bin} is that in Fig.~\ref{fig:velo-with-bin} the influence of binaries is taken into account, as described and quantified in Section~\ref{sec:binaries}.

It can be seen in Figs.~\ref{fig:velo-with-bin} and~\ref{fig:velo-no-bin} that the {\it observed} internal velocity dispersions of ETGs with luminosities $L_V \apprle 10^6 \, {\rm L}_{\odot}$ depend only weakly on the luminosity of the galaxies, if at all. Together with the assumptions of Newtonian dynamics and a common characteristic radius of 300 pc for all low-mass ETGs, this observation is the basis for the claim  by \citet{Strigari2008}, which is a common mass scale of about $10^7 \ {\rm M}_{\odot}$ for low-luminosity ETGs. \citet{Strigari2008}  suggest this mass scale as the characteristic mass of the least massive CDM-haloes, in which galaxies can form. The faintest ETGs would consist almost exclusively of CDM according to this interpretation. Using $R_{\rm e}=300$ pc and the data on $\sigma_{\rm obs}$ in equation~(\ref{eq:Mdyn}), this common mass scale of $10^7 {\rm M}_{\odot}$ for low-luminosity ETGs can be reproduced quite well, even though \citet{Strigari2008} obtained their result from a more involved modelling of the velocity dispersions in CDM-haloes.

However, observationally, the characteristic radii of faint ETGs do depend on their luminosities, if $R_{\rm e}$ is taken as an indicator of the characteristic radii (see, for instance, fig.~1 in \citealt{Misgeld2011}). This argues against the common mass scale for faint ETGs suggested by \citet{Strigari2008}, because according to them, the faint ETGs would mostly consist of dark matter even in their centres. The dark matter would therefore dominate the internal dynamics of the galaxies. The most straight-forward expectation is in this case that the distribution of the baryons reflects the distribution of the dark matter, i.e. extended dark matter haloes would tend to contain extended galaxies. Thus, if baryonic mass the DM-haloes contain does not depend on the total mass of the the DM-halo, the extension of the galaxies should not depend on their baryonic mass, and thus their luminosity. However, in reality, the least extended galaxies are also the ones that contain the least baryons. This would imply that the most concentrated dark matter haloes, which provide the deepest potential wells for a given halo mass, would nevertheless be the haloes that retain the least baryons, which is at least not intuitive. It is indeed problematic to reproduce the common mass-scale claimed by \citet{Strigari2008} in simulations \citep{Kroupa2010}.

Fig.~\ref{fig:velo-comp} shows a comparison between the median ratios of $\sigma_{\rm obs}/\sigma_{\rm pred}$, which are shown in the bottom panels of Figs.~\ref{fig:velo-no-bin} and~\ref{fig:velo-with-bin}. It is thus shown in Fig.~\ref{fig:velo-comp} how well $\sigma_{\rm obs}$ can be explained based on the probable amount of baryons in the ETGs, based on our assumptions regarding the laws of gravity governing the internal dynamics of ETGs (Newtonian Dynamics or Milgromian Dynamics), and our assumptions regarding the population of binaries in the ETGs. The dashed-dotted line at $\sigma_{\rm obs}/\sigma_{\rm pred}$ indicates equality between predictions and observations. If the relations for the typical values for $\sigma_{\rm obs}/\sigma_{\rm pred}$ are above that line, then $\sigma_{\rm obs}$ tends to be larger than $\sigma_{\rm pred}$. This is indeed the case for Newtonian dynamics at all luminosities, also when the effect of binaries on $\sigma_{\rm obs}$ is considered. Such a tendency also exists in Milgromian dynamics, but to a much lesser extent than in Newtonian dynamics. The conventional, but not the only way to interpret this finding is to assume that there is additional matter in the ETGs, which has not been accounted for in the estimates for $\sigma_{\rm pred}$ and $M_{\rm pred}$.

Fig.~\ref{fig:velo-comp} also illustrates that the impact of binaries on $\sigma_{\rm obs}$ is only noticeable below a certain luminosity. The reason is that $\sigma_0$ increases with mass (and thus luminosity) of the ETGs, while $\sigma_{\rm bin}$ is assumed to be the same in all of them, so that $\sigma_{\rm bin}$ has no appreciable impact on $\sigma_{\rm obs}$ for large $\sigma_0$ (cf. equation~\ref{eq:binaries}). In the case of Newtonian dynamics, binaries have virtually no effect on $\sigma_{\rm obs}$ of ETGs with $L_V \apprge 10^8 \, {\rm L}_{\odot}$, but a very strong effect on ETGs with $L_V \apprle 10^6 \, {\rm L}_{\odot}$. Milgromian dynamics boosts $\sigma_0$ especially in the faint ETGs, which have a low density. Thus, in the case of Milgromian dynamics, the effect of binaries on $\sigma_{\rm obs}$ is much milder, and is unlikely to be of practical relevance in ETGs with $L_V \apprge 10^5 \, {\rm L}_{\odot}$.

The best agreement between observations and predictions is achieved when Milgromian dynamics and a population of binaries are assumed for the ETGs. The median ratio between observed and predicted velocity dispersions rarely deviates by more than 30 per cent from unity. However, only for about 100 ETGs with luminosities between a few $10^8 \, {\rm L}_{\odot}$ and a few $10^9 \, {\rm L}_{\odot}$, the median ratio between observed and predicted velocity dispersions does not deviate significantly from unity. At other luminosities, the deviations may not be large, but they reflect properties of the data beyond purely statistical effects. The trend to a larger-than-expected $\sigma_{\rm obs}$ for ETGs with high luminosities is based on data on 500 ETGs. For the ETGs with low luminosities, the trend for $\sigma_{\rm obs}$ being above the expectation for ETGs with $L_V < 10^7 \, {\rm L}_{\odot}$ is based on only about 50 objects, but is undeniable nevertheless. If the ETGs are assumed to obey to Newtonian dynamics, or do not contain binaries, the modelled velocity dispersions underestimate the observed ones significantly more. On the other hand, the internal velocity dispersions of ETGs can neither in Milgromian dynamics nor in Newtonian dynamics fully be explained with our simple assumptions, even if every star is considered to be part a binary (i.e. assuming $f=1$ instead of $f=0.3$), while leaving the other parameters unchanged. Also then, the low-mass ETGs would still have internal velocity dispersions which tend to be higher than predicted, while the assumed population would contain much more and much tighter binaries than what seems consistent with observed stellar populations (see Section~\ref{sec:binaries} for a discussion of probable parameters for a population of binaries). This issue deserves more study though, because dynamical population synthesis predicts low-mass dwarf galaxies to have higher binary fractions than massive elliptical galaxies \citep{Marks2011}, and our approach to quantifying the effect of binaries is quite approximate.

\begin{figure*}
\centering
\includegraphics[scale=0.8]{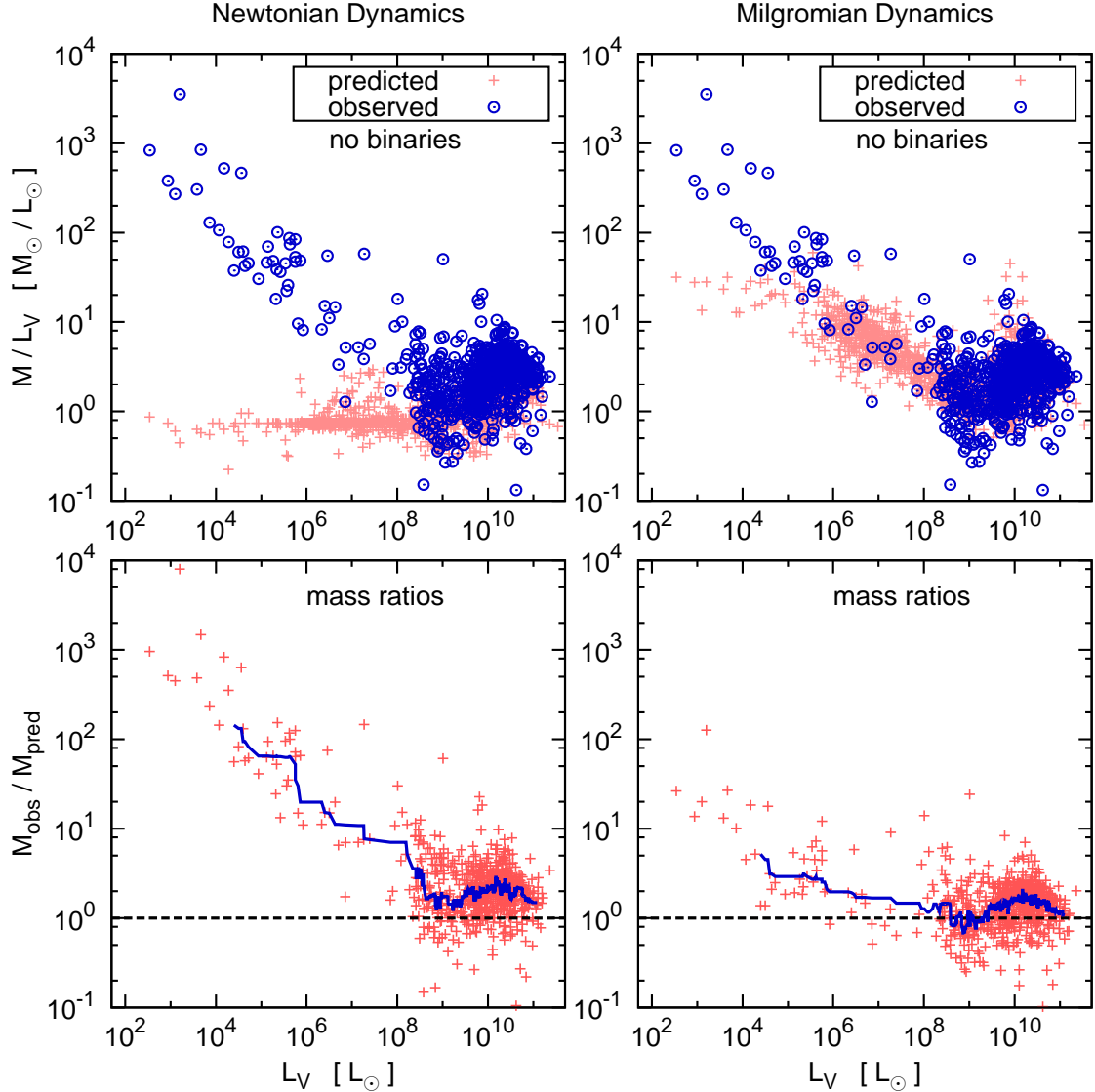}
\caption[Comparison between the predicted masses and $M_{\rm obs}$ of the ETGs, without considering a possible effect of binaries on the estimates.]{\label{fig:ML-no-bin} The Comparison between $M_{\rm pred}$ and $M_{\rm obs}$ of the ETGs, without considering a possible effect of binaries. In the panels on the left, Newtonian dynamics is assumed for calculating $M_{\rm pred}$, and Milgromian dynamics in the panels on the right. In the two top panels, the values for $M_{\rm obs}$ are plotted on top of the values for $M_{\rm pred}$. The two bottom panels show the ratio between $M_{\rm obs}$ and $M_{\rm pred}$ for the ETGs where both quantities are known. The solid lines in the bottom panels quantifies the change of the median value of $M_{\rm obs}/M_{\rm pred}$ with luminosity. Note that the method for finding these median values does not allow us to assign median values to the luminosities of the ten faintest and the ten brightest galaxies shown in the figure. The method is described in detail in Section~\ref{sec:velocities}. The dashed lines in these panels indicate equality between the predictions and the observations.}
\end{figure*}

\begin{figure*}
\centering
\includegraphics[scale=0.8]{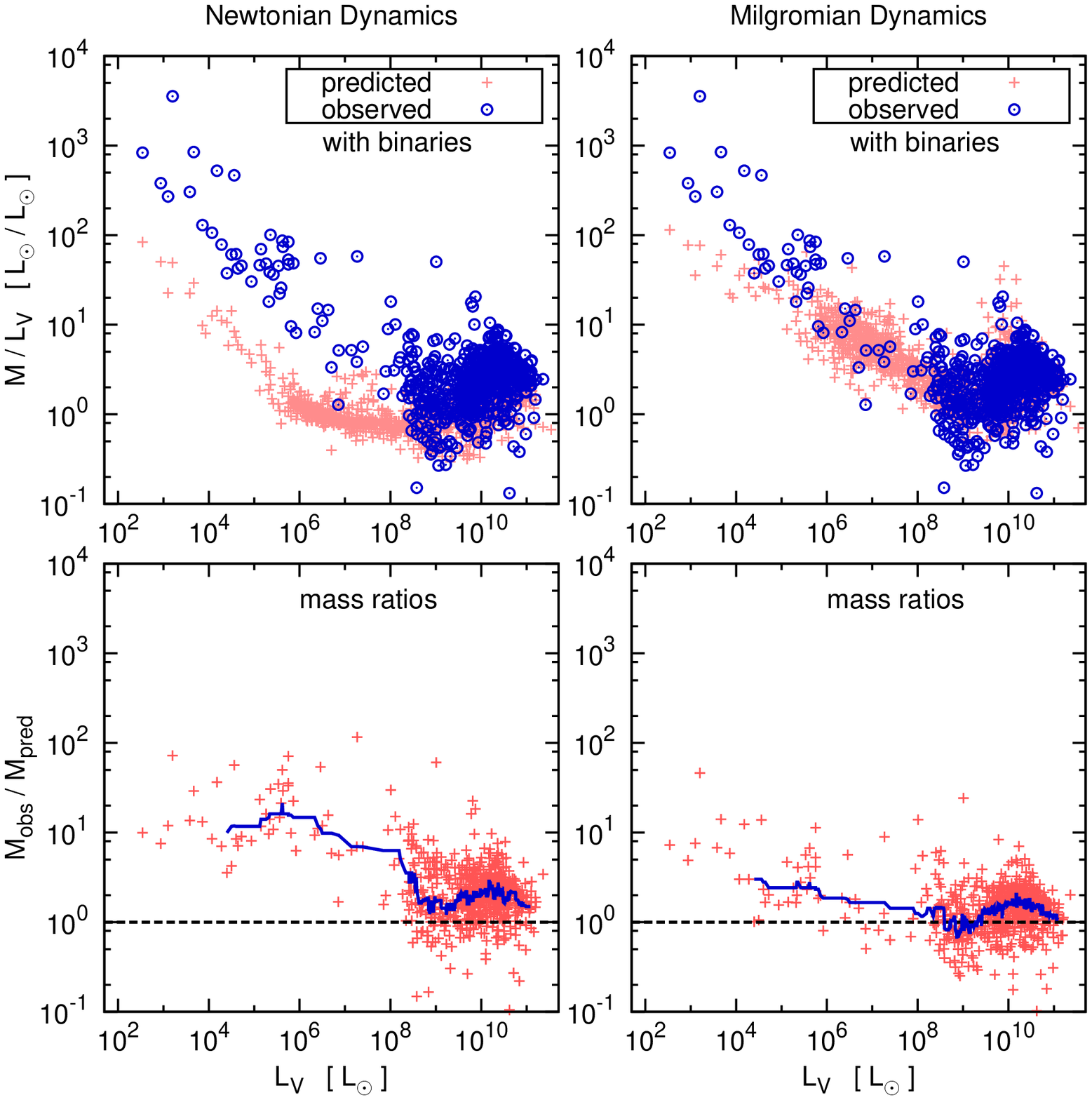}
\caption[Comparison between the predicted masses and $M_{\rm obs}$ of the ETGs, considering a possible effect of binaries on the observed velocity dispersions.]{\label{fig:ML-with-bin} As Fig.~\ref{fig:ML-no-bin}, but assuming that 30 per cent of the stars in the ETGs are part of a binary.}
\end{figure*}

\begin{figure*}
\includegraphics[scale=0.8]{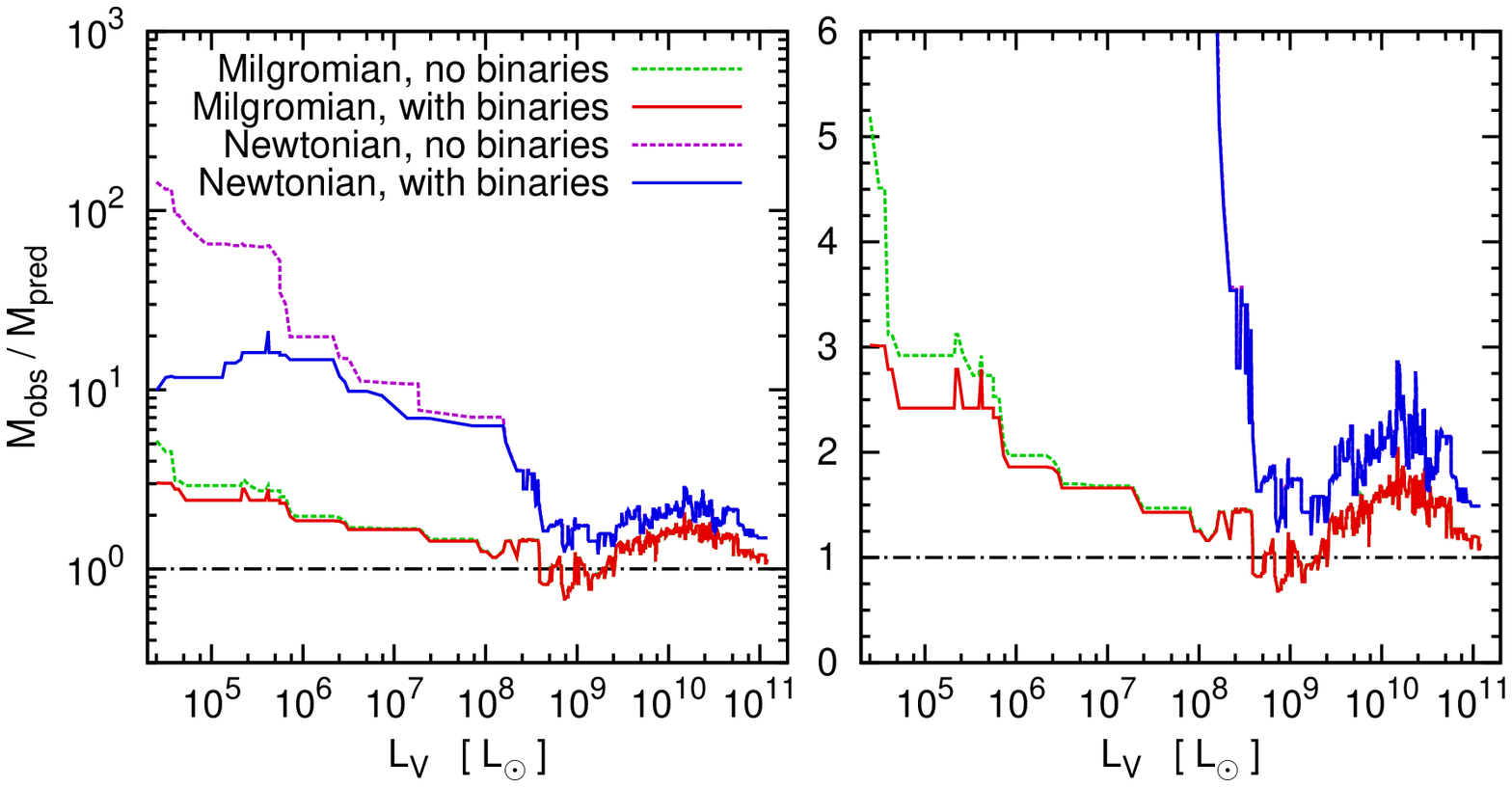}
\centering
\caption[Comparison between the medians of the ratios between $M_{\rm obs}$ and the predicted velocity dispersions of ETGs.]{\label{fig:ML-comp} Comparison between the medians of the ratios between $M_{\rm obs}$ and the predicted masses of ETGs, as shown in the bottom panels of Figs.~\ref{fig:ML-no-bin} and~\ref{fig:ML-with-bin}. The left panel shows the full range of the relations quantifying the ratios between $M_{\rm obs}$ and the predicted masses of the ETGs. The assumptions made for the different predictions on the masses of the ETGs are indicated in the left panel of the figure. The right panel of this figure shows the same relations as the left panel, but zooms in on the range where the ratio between observed and predicted masses is close to unity.}
\end{figure*}

\subsection[Mass-to-light ratios and masses]{Mass-to-light ratios and masses}
\label{sec:MLratios}

Figs.~\ref{fig:ML-no-bin} and~\ref{fig:ML-with-bin} show comparisons between the predictions for the masses and mass-to-light ratios of ETGs.

The different predictions for the masses, $M_{\rm pred}$, shown in Figs.~\ref{fig:ML-no-bin} and~\ref{fig:ML-with-bin} are obtained with equation~(\ref{eq:Mdyn}), using $\sigma_{\rm pred}$ (i.e. the different predictions for $\sigma_{\rm obs}$), and $\sigma_{\rm obs}$ itself, respectively. As noted in Section~\ref{sec:masses}, the resulting $M_{\rm pred}$ do not necessarily represent masses, in the sense that they actually measure the amount of matter in the galaxy. Instead, $M_{\rm N}^{\rm bin}$,  $M_{\rm M}$ and $M_{\rm M}^{\rm bin}$ are quantities that help to determine how well $M_{\rm obs}$ can be understood in terms of the assumptions that enter the estimates of $M_{\rm N}^{\rm bin}$,  $M_{\rm M}$, and $M_{\rm M}^{\rm bin}$, respectively. ($M_{\rm N}$ on the other hand is a real mass estimate, while the assumptions that enter it and the results obtained from it can be wrong.) $M_{\rm obs}$ is usually interpreted as an estimate for the actual mass of an ETG, but depends on assumptions itself, notably that Newtonian dynamics is valid in the ETGs, that the ETG is in virial equilibrium and that the effect of binaries on $\sigma_{\rm obs}$ is negligible. However, in contrast to  $M_{\rm N}^{\rm bin}$,  $M_{\rm M}$ and $M_{\rm M}^{\rm bin}$, $M_{\rm obs}$, as well as $M_{\rm N}$, truly measure the amount of matter in the ETG, if the assumptions made for estimating $M_{\rm obs}$ or $M_{\rm N}$ are correct. If $M_{\rm obs}>M_{\rm N}$, $M_{\rm obs}>M_{\rm N}^{\rm bin}$, $M_{\rm obs}>M_{\rm M}$, or $M_{\rm obs}>M_{\rm M}^{\rm bin}$ is found for an ETG, one way to interpret such a result is that the galaxy contains additional matter, while other interpretations are possible as well (see Section~\ref{sec:conclusion}).

In Fig.~\ref{fig:ML-no-bin}, $M_{\rm obs}$ is compared to $M_{\rm N}$ and $M_{\rm M}$, respectively. Thus a possible effect of binaries on the mass estimates is neglected in this figure.

In the two top panels of Fig.~\ref{fig:ML-no-bin}, the estimates for $M_{\rm obs}/L_V$ are plotted on top of $M_{\rm N}/L_V$ (left panel) and $M_{\rm M}/L_V$ (right panel). In particular for ETGs with $V$-band luminosities $10^5 \, {\rm L}_{\odot} \apprle L_V \apprle 10^9 \, {\rm L}_{\odot}$, $M_{\rm obs}$ cannot be calculated for all ETG for which $M_{\rm N}$ and $M_{\rm M}$ can be calculated. The reason is that $\sigma_{\rm obs}$ is not yet known for many of these faint ETGs, since observations that allow to derive their $\sigma_0$ become the more challenging, the fainter they are and the lower their surface brightness is. Thus, $\sigma_0$ has been estimated for many faint ETGs in the Local Group by now, but the majority of ETGs with $V$-band luminosities $10^5 \, {\rm L}_{\odot} \apprle L_V \apprle 10^9 \, {\rm L}_{\odot}$ shown in Fig.~\ref{fig:ML-no-bin} are members of galaxy clusters other than the Local Group. For them, the effort of estimating their $\sigma_0$ would require extremely deep and time-consuming observations, if they are possible at all with presently available instruments.

For all ETGs for which $M_{\rm obs}$, $M_{\rm N}$ and $M_{\rm M}$ are available, $M_{\rm obs}/M_{\rm N}$ is shown in the bottom left panel and $M_{\rm obs}/M_{\rm M}$ is shown in the bottom right panel. The solid (blue) lines in both panels quantify the median of the ratios $M_{\rm obs}/M_{\rm N}$, and $M_{\rm obs}/M_{\rm M}$, respectively, as a function of $L_V$. These relations are obtained as described in Section~\ref{sec:velocities} for the median of $\sigma_{\rm obs}/\sigma_{\rm N}$.

Fig.~\ref{fig:ML-with-bin} is organised as Fig.~\ref{fig:ML-no-bin}, but in Fig.~\ref{fig:ML-with-bin}, $M_{\rm N}$ is replaced by $M_{\rm N}^{\rm bin}$, and $M_{\rm M}$ is replaced by $M_{\rm M}^{\rm bin}$. Thus, the difference between the two figures is that in Fig.~\ref{fig:ML-with-bin}, the influence of binaries is taken into account in the predictions on the masses of ETGs.

Fig.~\ref{fig:ML-comp} shows a comparison between the median ratios of $M_{\rm obs}/M_{\rm pred}$, as shown in the bottom panels of Figs.~\ref{fig:ML-no-bin} and~\ref{fig:ML-with-bin}. It is thus shown in Fig.~\ref{fig:ML-comp} how well $M_{\rm obs}$ can be explained based on the probable amount of baryons in the ETGs, based on our assumptions regarding the laws of gravity governing the internal dynamics of ETGs (Newtonian Dynamics or Milgromian Dynamics), and our assumptions regarding the population of binaries in the ETGs.

Basically, Fig.~\ref{fig:ML-comp} shows the same trends as Fig.~\ref{fig:velo-comp}. However, with Fig.~\ref{fig:ML-comp}, it is more straight forward to discuss what these trends imply for mass estimates and the matter content of ETGs. This is done in the following section.

\section{Discussion and Conclusion}
\label{sec:conclusion}

Motivated by the notion that many, if not most dwarf elliptical galaxies might be tidal dwarf galaxies \citep{Okazaki2000,Kroupa2010,Dabringhausen2013}, and would then contain basically no dark matter (cf. \citealt{Barnes1992a,Bournaud2010}), we discuss how the elevated internal velocity dispersions of ETGs could be understood without dark matter. In this context, we have made two assumptions, which are well motivated by observations. The first assumption is that the ETGs obey to Milgromian dynamics \citep{Milgrom1983a} instead of Newtonian Dynamics. Milgromian dynamics formulates a deviation from Newtonian gravity in the limit of very weak gravitational fields, which is extremely successful in quantifying the dynamics of galaxies \citep{Famaey2012}. The second assumption is that many stars are part of a binary system, as it is observed in resolved stellar populations \citep{Duquennoy1991,Carney2005}, including some dwarf elliptical galaxies \citep{Simon2011,Minor2013}.

It is well known that if virial equilibrium and Newtonian dynamics are assumed, the observed velocity dispersions, $\sigma_{\rm obs}$, of ETGs tend to be larger than what is expected based on the baryonic matter in ETGs. This is true over the whole luminosity range of ETGs, even though the typical magnitude of the discrepancy depends on the luminosity of the ETGs (see Figs.~\ref{fig:velo-no-bin} to~\ref{fig:velo-comp}). Therefore, masses estimated based on $\sigma_{\rm obs}$ are larger than expected as well (see Figs.~\ref{fig:ML-no-bin} to~\ref{fig:ML-comp}). This discrepancy is probably even larger than apparent in the present paper, since isotropy is assumed here for the stellar orbits in the ETGs (see the forth assumption in Section~\ref{sec:assumptions}), while the observed LOS velocity distributions of over 2000 ETGs studied by \citet{Vudragovic2016} indicate that radial orbits are preferred over tangential orbits in them. Consequently, \citet{Vudragovic2016} find for their sample of ETGs that more realistic mass estimates based on the internal dynamics are on average about 10 per cent higher than expected under the assumption of isotropy. The noted discrepancy between mass estimates based on the internal dynamics of ETGs and mass estimates based on the amount of baryons detected in them is usually interpreted as a presence of non-baryonic dark matter, especially in massive systems which are unlikely to be disturbed by tidal forces.

As it is expected, the discrepancy between observed and predicted velocity dispersions (and consequently the discrepancy on the mass estimates based on them) decreases if Milgromian dynamics instead of Newtonian dynamics is assumed. However, also in this case, there are luminosity intervals, where $\sigma_{\rm obs}$ is on average higher than values predicted based on the baryonic matter in the ETGs. This concerns objects more luminous than a few $10^9 \ {\rm L}_{\odot}$ and objects less luminous than $10^6 \ {\rm L}_{\odot}$.

In Newtonian dynamics, binaries could significantly boost the observed velocity dispersions of systems with luminosities $L_V \apprle 10^6 {\rm L}_{\odot}$, however not to an extent that the dynamics of the systems could be explained without dark matter, as long as they are in virial equilibrium. This does not exclude that the $\sigma_{\rm obs}$ of individual low-mass ETGs could be explained within the measurement uncertainties, if binaries are considered. However, the trend to strongly elevated $\sigma_{\rm obs}$ for the population of low-mass ETGs as a whole remains obvious \citep{Hargreaves1996,McConnachie2010}.

In Milgromian dynamics, the expected velocity dispersions of centre-of-mass systems in the gravitational potential of a galaxy are increased compared to the case of Newtonian dynamics. The contribution of binary stars to $\sigma_{\rm obs}$ is however independent of the motions of the stars in the potential of the galaxy. In consequence, the contribution of binary stars to $\sigma_{\rm obs}$ in Milgromian dynamics is small in comparison to the case of Newtonian dynamics; also because typical binary systems are so tightly bound that they can still be considered as Newtonian systems, so that assuming Milgromian dynamics has no effect on the orbital velocities of typical binaries. In consequence, the effect of binaries is rather weak on the $\sigma_{\rm obs}$ of ETGs with luminosities $L_V \apprle 10^5 {\rm L}_{\odot} $, and basically non-existent for more luminous ETGs. In Newtonian dynamics, the effect of binaries on $\sigma_{\rm obs}$ of low-luminosity ETGs is much more pronounced, and includes ETGs with luminosities $L_V \apprle10^6 {\rm L}_{\odot}$ (see Fig.~\ref{fig:velo-comp}).

Thus, also in Milgromian dynamics, binaries cannot fully explain the high $\sigma_{\rm obs}$ of ETGs. This is even the case when the observationally motivated binary fraction $f=0.3$ is increased to $f=1.0$, while leaving the other parameters that describe the assumed binary population unchanged. In consequence, the observed $\sigma_{\rm obs}$, and consequently the $M_{\rm obs}$, tend to be elevated for ETGs with luminosities $L_V \apprle10^6 {\rm L}_{\odot}$ and for ETGs with luminosities $L_V \apprge10^{10} {\rm L}_{\odot}$, even if Milgromian dynamics is assumed. We note that \citet{Janz2016} recently used the dynamical models of fast-rotator ETGs by \citet{Cappellari2015} based on ATLAS$^{\rm 3D}$ and SLUGGS data to show that Milgromian predictions based on the interpolating function used in our present study (the 'simple' $\mu$-function; equation~\ref{eq:mu}) are globally well in-line with the data. However, they note that spiral galaxies are better fitted with the 'standard' interpolating function (equation~10 in \citealt{Milgrom1983b}). Nevertheless, we insist on the fact that also the 'simple' $\mu$-function has been shown in the past to reproduce brilliantly the rotation curves of spiral galaxies \citep{Gentile2011}. The quality of the fits is in this case more a matter of the adopted stellar $M/L$-ratios than of the adopted $\mu$-function, and the $M/L$-ratios of stellar populations depend on their age, their metallicity and their IMF, which are all uncertain quantities for unresolved stellar populations.

However, apart from the inclusion of binaries, there are also two other assumptions made in the present study that merit some discussion.

The first assumption is that the ETGs are in virial equilibrium. If this assumption is wrong, the mass of the system will be overestimated (see e.g. \citealt{Kroupa1997}).

The second assumption is that the ETGs formed with a canonical IMF as it is parametrised in \citet{Kroupa2001} or \citet{Chabrier2003}, i.e. an IMF that has the Salpeter-slope for stellar masses larger than $0.5 \ {\rm M}_{\odot}$, but flattens for lower stellar masses\footnote{Within observational limitations, both these formulations of the canonical IMF are identical \citep{Kroupa2013}}. This assumption is implicitly made by adopting the values for the $M_{\rm s}$ of the ETGs taken from \citet{Dabringhausen2016}. ETGs are stellar systems with rather old stellar populations, and in such systems there are in principle two possibilities how a variation of the IMF could increase the mass-to-light ratio with respect to the canonical IMF. The first possibility is an IMF overabundant in low-mass stars (bottom-heavy IMF), so that more mass is attributed to the total stellar population by a large population of very faint low-mass stars. The second possibility is an IMF overabundant in high mass stars, which have evolved into stellar remnants. Thus, for a given luminosity, either a bottom-heavy IMF or a top-heavy IMF would imply a higher mass for the ETG in comparison to an ETG that formed with the canonical IMF\citep{Cappellari2012}.

Concerning the ETGs with luminosities $L_V \apprge10^{10} {\rm L}_{\odot}$, we note that their masses and densities imply that they can hardly be disturbed by tidal fields. Thus, the assumption of virial equilibrium is well justified for them. For their IMFs, we note that \citet{Tortora2014} already discussed the observed $\sigma_{\rm obs}$ and $M_{\rm obs}$ in the inner parts of 220 massive ETGs covered in the ATLAS$^{\rm 3D}$-survey \citep{Cappellari2011} under the premise of Milgromian dynamics instead of non-baryonic dark matter. They found that also in Milgromian dynamics, the observed $\sigma_{\rm obs}$ and $M_{\rm obs}$ are better explained with a Salpeter IMF, i.e. an IMF that has (in contrast to the canonical IMF) the Salpeter-slope down to the lowest stellar masses. \citet{Samurovic2014} confirms this trend in a very detailed study of 10 massive ETGs. However, since the canonical IMF has often been argued to be universal for all star formation based on observed stellar populations in the Local Universe, indications for a Salpeter IMF in massive ETGs and a canonical IMF in nearby star clusters contradict the paradigm of an universal IMF\citep{Kroupa2013}. On the other hand, the notion of a canonical IMF in the most massive ETGs has indeed been challenged in the past years, not only based on their mass-to-light ratios. Some absorption lines in spectra of galaxies are sensitive to the presence of low-mass stars, and the observed strength of these spectral features in the most luminous ETGs have been taken as evidence that the IMF in them was highly bottom-heavy when most of their stars formed \citep{vanDokkum2010,Ferreras2013}. However, the high mass-to-light ratios that the results by \citet{vanDokkum2010} imply may be in conflict with constrains on the mass-to-light ratios of elliptical galaxies set by mass estimates from gravitational lensing \citet{Leier2016}. \citet{Weidner2013}, in contrast, argue that the galaxy-wide IMF in the most massive ETGs was top-heavy when most of their stars formed. They base their argument on the notions that the IMF may be top-heavy in the most massive star clusters \citep{Dabringhausen2012,Marks2012b}, that massive star clusters form if its star formation rate in a galaxy is high \citep{Weidner2004b}, and that the star formation rates in massive ETGs were high when they formed \citep{Thomas2005}. However, using the fact that neutron stars become visible as X-ray sources when they accrete matter, \citet{Peacock2014} searched for a dependency between the number of neutron stars per unit luminosity and the total luminosity of ETGs. Such a dependency would be expected, if the IMF for high-mass stars varies with the luminosity of ETGs, but \citet{Peacock2014} found that their results are consistent with a universal (i.e. non-varying, but necessarily canonical) IMF in ETGs. However, \citet{Peacock2014} did not specifically test a galaxy-wide top-heavy IMF, Moreover, \citet{Peacock2014} implicitly assume Newtonian dynamics in their study and a more moderate variation of the IMF than the case they discuss is not excluded. Thus, in summary, The discrepancy between $M_{\rm pred}$ and $M_{\rm obs}$ lies well within the range of what could be explained with a variation of the IMF. However, the nature of the unidentified matter in centres of high-luminosity ETGs is still an open question, but it is not necessarily non-baryonic.

Concerning the ETGs with luminosities $L_V \apprle 10^6 \ {\rm L}_{\odot}$, i.e. the low-mass ETGs where $\sigma_{\rm obs}$ and consequently $M_{\rm obs}$ is also in Milgromian dynamics clearly above $M_{\rm pred}$, we note that the Leo~T dSph may contain more gas than stars \citep{Belokurov2007}, but usually dSphs contain only little baryonic matter apart from the baryonic matter in stars. Also a variation of the IMF could hardly alter the mass of the stellar systems to the extent that Figs.~\ref{fig:ML-no-bin} to~\ref{fig:ML-comp} imply for low-mass galaxies. However, at least in Newtonian dynamics, low-mass, dark-matter-free systems in the Milky Way are vulnerable to tidal forces, which can increase the internal velocity dispersions, and consequently elevate the estimates for the mass of the object under the (in this case wrong) assumption that they are in tidal equilibrium (\citealt{Kroupa1997,Casas2012}, Dom\'{i}nguez et al., submitted). Tidal forces may even disrupt a stellar system. Such a system may still be detected as an over-density of stars, for which $\sigma_{\rm obs}$ and $R_{\rm e}$ can be observed, but an estimate of the mass of the system based on equation~(\ref{eq:Mdyn}), where virial equilibrium is assumed, is clearly impossible. If such an estimate is performed nevertheless, the result is an extreme over-estimate of the mass of the object \citep{Kroupa1997}.

Besides their (at least apparently) high mass-to-light ratios, there is also other evidence for low-mass ETGs being tidally disturbed. \citet{McGaugh2010} found that the ellipticity of dwarf galaxies in the Local Group is correlated with their luminosity, so that the least luminous ETGs (i.e. the most vulnerable to tidal forces) tend to be the least spherical, and a  departure from a round shape can be caused by external forces acting on a galaxy. Furthermore, \citet{McGaugh2010} find that the ETGs that are the closest to their hosts (the Milky Way or M31), i.e the ones that are subjected to the strongest tidal forces, also tend to be the ones that are the least spherical. More detailed observations of single galaxies show further peculiarities that can be interpreted as evidence for non-equilibrium dynamics for many of these ETGs, like elongated or irregular shapes, assymetic surface- brightness profiles, or asymmetric velocity-dispersion profiles (e.g. \citealt{Belokurov2006,Belokurov2007,Walker2009,Munoz2010,Willman2011,Deason2012}). We note, however, that most if not all of these peculiarities can also be explained, if their stellar populations formed as groups of star clusters embedded in dark-matter haloes \citep{Assmann2013a,Assmann2013b}. On the other hand, substructures are more likely to be preserved in cored dark matter haloes, which are motivated by observational properties of low-mass ETGs \citep{Gilmore2007}, but are still a huge challenge to $\Lambda$CDM-theory, especially in low-mass galaxies \citep{Governato2012,diCintio2014,Onorbe2015}.

Thus, the internal velocity dispersions and the internal structure of low-mass ETGs can in principle be understood as consequences of their formation in dark-matter haloes, while for an explanation without dark matter non-equilibrium dynamics in low-mass objects has to be assumed. There is however strong evidence that many, if not most low-mass ETGs are tidal dwarf galaxies (e.g. \citealt{Kroupa2010,Dabringhausen2013}) and as such basically free of dark matter (e.g. \citealt{Bournaud2010}). For instance, it is apparently quite common for low-mass ETGs to be part of rotationally supported disks of satellites around more massive galaxies (e.g. \citealt{Ibata2014}). This could easily be understood by the conservation of angular momentum in the encounters of primordial galaxies if the low-mass ETGs are TDGs, while the $\Lambda$CDM-model predicts a much more isotropic distribution and predominately random motions for primordial dwarf galaxies around their hosts \citep{Kroupa2005,Pawlowski2014}.

In the context of understanding ETGs (especially low-mass ETGs) as objects without non-baryonic dark matter, Milgromian dynamics makes the discrepancies between observed and predicted internal dynamics either disappear (at least for objects with luminosities close to $10^9 \ {\rm L}_{\odot}$) or become less extreme and therefore easier to explain. For high-mass ETGs, assuming Milgromian dynamics implies that the discrepancy between observed and predicted internal velocity dispersions can be explained with a galaxy-wide IMF that varies within reasonable parameters. For low-mass ETGs, assuming Milgromian dynamics implies that some of them are probably disturbed by tidal fields, but not necessarily all of them, as can be seen in Figs.~\ref{fig:velo-no-bin} to~\ref{fig:velo-comp}. This scenario appears more likely than the case that virtually {\it all} low-mass ETGs are out of virial equilibrium, as assuming Newtonian dynamics would suggest for them, unless they contain non-baryonic dark matter.

It remains to be seen how effective the creation of TDGs in Milgromian dynamics is, and how they evolve in the tidal fields in Milgromian dynamics, and thus how well the notion that low-mass ETGs are TDGs that obey Milgromian dynamics agrees with the observations. \citet{Candlish2015} and \citet{Lueghausen2015} have implemented Milgromian dynamics into the RAMSES code \citep{Teyssier2002}, and have thereby created tools that will help to tackle such questions numerically. Ultimately, understanding the nature of low-mass ETGs, i.e. whether they are primordial objects that formed in dark matter haloes or whether they are TDGs, may help to create a revised cosmological model, since the doubts have been cast on the $\Lambda$CDM-model, which is the currently prevailing cosmological model \citep{Kroupa2010,Famaey2012,Kroupa2012,Famaey2013,Kroupa2015}.

\section*{Acknowlegdements}
We wish to thank the referee, Srdjan Samurovi\'{c}, for helpful comments that improved the paper. JD gratefully acknowledges a stipend from the University of Bonn during the final phase of his PhD and funding through FONDECYT concurso postdoctorado grant~3140146. MF is funded by FONDECYT~1130521 and Basal-CATA. Throughout the preparation of this paper, we have made extensive use of NASA's Astrophysics Data System. We also acknowledge usage of the NASA/IPAC Extragalactic Database. 

\bibliographystyle{mn2e}
\bibliography{a-e,f-j,i-m,n-z}

\begin{thebibliography}{}

\bibitem[\protect\citeauthoryear{{Abt} \& {Levy}}{{Abt} \&
  {Levy}}{1976}]{Abt1976}
{Abt} H.~A.,  {Levy} S.~G.,  1976, ApJS, 30, 273

\bibitem[\protect\citeauthoryear{{Angus}, {Famaey} \& {Buote}}{{Angus}
  et~al.}{2008}]{Angus2008}
{Angus} G.~W.,  {Famaey} B.,    {Buote} D.~A.,  2008, MNRAS, 387, 1470

\bibitem[\protect\citeauthoryear{{Assmann}, {Fellhauer}, {Wilkinson} \&
  {Smith}}{{Assmann} et~al.}{2013a}]{Assmann2013a}
{Assmann} P.,  {Fellhauer} M.,  {Wilkinson} M.~I.,    {Smith} R.,  2013, MNRAS,
  432, 274

\bibitem[\protect\citeauthoryear{{Assmann}, {Fellhauer}, {Wilkinson}, {Smith}
  \& {Bla{\~n}a}}{{Assmann} et~al.}{2013b}]{Assmann2013b}
{Assmann} P.,  {Fellhauer} M.,  {Wilkinson} M.~I.,  {Smith} R.,    {Bla{\~n}a}
  M.,  2013, MNRAS, 435, 2391

\bibitem[\protect\citeauthoryear{{Barnes} \& {Hernquist}}{{Barnes} \&
  {Hernquist}}{1992}]{Barnes1992a}
{Barnes} J.~E.,  {Hernquist} L.,  1992, Nature, 360, 715

\bibitem[\protect\citeauthoryear{{Belokurov~et~al.}}{{Belokurov~et~al.}}{2006}]{Belokurov2006}
{Belokurov~et~al.} 2006, ApJ, 647, L111

\bibitem[\protect\citeauthoryear{{Belokurov~et~al.}}{{Belokurov~et~al.}}{2007}]{Belokurov2007}
{Belokurov~et~al.} 2007, ApJ, 654, 897

\bibitem[\protect\citeauthoryear{{Bertin}, {Ciotti} \& {Del Principe}}{{Bertin}
  et~al.}{2002}]{Bertin2002}
{Bertin} G.,  {Ciotti} L.,    {Del Principe} M.,  2002, A\&A, 386, 149

\bibitem[\protect\citeauthoryear{{Binney} \& {Tremaine}}{{Binney} \&
  {Tremaine}}{1987}]{Binney1987}
{Binney} J.,  {Tremaine} S.,  1987, {Galactic dynamics}.
Princeton University Press

\bibitem[\protect\citeauthoryear{{Bournaud}}{{Bournaud}}{2010}]{Bournaud2010}
{Bournaud} F.,  2010, Advances in Astronomy, 2010

\bibitem[\protect\citeauthoryear{{Bournaud} \& {Duc}}{{Bournaud} \&
  {Duc}}{2006}]{Bournaud2006}
{Bournaud} F.,  {Duc} P.-A.,  2006, A\&A, 456, 481

\bibitem[\protect\citeauthoryear{{Bournaud}, {Duc}, {Brinks}, {Boquien},
  {Amram}, {Lisenfeld}, {Koribalski}, {Walter} \& {Charmandaris}}{{Bournaud}
  et~al.}{2007}]{Bournaud2007}
{Bournaud} F.,  {Duc} P.-A.,  {Brinks} E.,  {Boquien} M.,  {Amram} P.,
  {Lisenfeld} U.,  {Koribalski} B.~S.,  {Walter} F.,    {Charmandaris} V.,
  2007, Science, 316, 1166

\bibitem[\protect\citeauthoryear{{Bournaud}, {Duc} \& {Emsellem}}{{Bournaud}
  et~al.}{2008}]{Bournaud2008}
{Bournaud} F.,  {Duc} P.-A.,    {Emsellem} E.,  2008, MNRAS, 389, L8

\bibitem[\protect\citeauthoryear{{Bruzual} \& {Charlot}}{{Bruzual} \&
  {Charlot}}{2003}]{Bruzual2003}
{Bruzual} G.,  {Charlot} S.,  2003, MNRAS, 344, 1000

\bibitem[\protect\citeauthoryear{{Candlish}, {Smith} \& {Fellhauer}}{{Candlish}
  et~al.}{2015}]{Candlish2015}
{Candlish} G.~N.,  {Smith} R.,    {Fellhauer} M.,  2015, MNRAS, 446, 1060

\bibitem[\protect\citeauthoryear{{Caon}, {Capaccioli} \& {D'Onofrio}}{{Caon}
  et~al.}{1993}]{Caon1993}
{Caon} N.,  {Capaccioli} M.,    {D'Onofrio} M.,  1993, MNRAS, 265, 1013

\bibitem[\protect\citeauthoryear{{Cappellari~et~al.}}{{Cappellari~et~al.}}{2011}]{Cappellari2011}
{Cappellari~et~al.} 2011, MNRAS, 413, 813

\bibitem[\protect\citeauthoryear{{Cappellari~et~al.}}{{Cappellari~et~al.}}{2012}]{Cappellari2012}
{Cappellari~et~al.} 2012, Nature, 484, 485

\bibitem[\protect\citeauthoryear{{Cappellari~et~al.}}{{Cappellari~et~al.}}{2013}]{Cappellari2013}
{Cappellari~et~al.} 2013, MNRAS, 432, 1709

\bibitem[\protect\citeauthoryear{{Cappellari~et~al.}}{{Cappellari~et~al.}}{2015}]{Cappellari2015}
{Cappellari~et~al.} 2015, ApJ, 804, L21

\bibitem[\protect\citeauthoryear{{Carney}, {Aguilar}, {Latham} \&
  {Laird}}{{Carney} et~al.}{2005}]{Carney2005}
{Carney} B.~W.,  {Aguilar} L.~A.,  {Latham} D.~W.,    {Laird} J.~B.,  2005, AJ,
  129, 1886

\bibitem[\protect\citeauthoryear{{Casas}, {Arias}, {Pe{\~n}a Ram{\'{\i}}rez} \&
  {Kroupa}}{{Casas} et~al.}{2012}]{Casas2012}
{Casas} R.~A.,  {Arias} V.,  {Pe{\~n}a Ram{\'{\i}}rez} K.,    {Kroupa} P.,
  2012, MNRAS, 424, 1941

\bibitem[\protect\citeauthoryear{{Chabrier}}{{Chabrier}}{2003}]{Chabrier2003}
{Chabrier} G.,  2003, PASP, 115, 763

\bibitem[\protect\citeauthoryear{{Dabringhausen} \&
  {Fellhauer}}{{Dabringhausen} \& {Fellhauer}}{2016}]{Dabringhausen2016}
{Dabringhausen} J.,  {Fellhauer} M.,  2016, MNRAS, 460, 4492

\bibitem[\protect\citeauthoryear{{Dabringhausen} \& {Kroupa}}{{Dabringhausen}
  \& {Kroupa}}{2013}]{Dabringhausen2013}
{Dabringhausen} J.,  {Kroupa} P.,  2013, MNRAS, 429, 1858

\bibitem[\protect\citeauthoryear{{Dabringhausen}, {Kroupa}, {Pflamm-Altenburg}
  \& {Mieske}}{{Dabringhausen} et~al.}{2012}]{Dabringhausen2012}
{Dabringhausen} J.,  {Kroupa} P.,  {Pflamm-Altenburg} J.,    {Mieske} S.,
  2012, ApJ, 747, 72

\bibitem[\protect\citeauthoryear{{Dariush~et~al.}}{{Dariush~et~al.}}{2016}]{Dariush2016}
{Dariush~et~al.} 2016, MNRAS, 456, 2221

\bibitem[\protect\citeauthoryear{{Deason}, {Belokurov}, {Evans}, {Watkins} \&
  {Fellhauer}}{{Deason} et~al.}{2012}]{Deason2012}
{Deason} A.~J.,  {Belokurov} V.,  {Evans} N.~W.,  {Watkins} L.~L.,
  {Fellhauer} M.,  2012, MNRAS, 425, L101

\bibitem[\protect\citeauthoryear{{Di Cintio}, {Brook}, {Dutton}, {Macci{\`o}},
  {Stinson} \& {Knebe}}{{Di Cintio} et~al.}{2014}]{diCintio2014}
{Di Cintio} A.,  {Brook} C.~B.,  {Dutton} A.~A.,  {Macci{\`o}} A.~V.,
  {Stinson} G.~S.,    {Knebe} A.,  2014, MNRAS, 441, 2986

\bibitem[\protect\citeauthoryear{{Dominguez}, {Fellhauer}, {Bla{\~n}a},
  {Farias}, {Dabringhausen}, {Candlish}, {Smith} \& {Choque}}{{Dominguez}
  et~al.}{2016}]{Dominguez2016}
{Dominguez} R.,  {Fellhauer} M.,  {Bla{\~n}a} M.,  {Farias} J.-P.,
  {Dabringhausen} J.,  {Candlish} G.~N.,  {Smith} R.,    {Choque} N.,  2016,
  ArXiv e-prints

\bibitem[\protect\citeauthoryear{{Duc}, {Bournaud} \& {Masset}}{{Duc}
  et~al.}{2004}]{Duc2004}
{Duc} P.-A.,  {Bournaud} F.,    {Masset} F.,  2004, A\&A, 427, 803

\bibitem[\protect\citeauthoryear{{Duc} \& {Mirabel}}{{Duc} \&
  {Mirabel}}{1994}]{Duc1994}
{Duc} P.-A.,  {Mirabel} I.~F.,  1994, A\&A, 289, 83

\bibitem[\protect\citeauthoryear{{Duc} \& {Mirabel}}{{Duc} \&
  {Mirabel}}{1998}]{Duc1998}
{Duc} P.-A.,  {Mirabel} I.~F.,  1998, A\&A, 333, 813

\bibitem[\protect\citeauthoryear{{Duc}, {Paudel}, {McDermid}, {Cuillandre},
  {Serra}, {Bournaud}, {Cappellari} \& {Emsellem}}{{Duc}
  et~al.}{2014}]{Duc2014}
{Duc} P.-A.,  {Paudel} S.,  {McDermid} R.~M.,  {Cuillandre} J.-C.,  {Serra} P.,
   {Bournaud} F.,  {Cappellari} M.,    {Emsellem} E.,  2014, MNRAS, 440, 1458

\bibitem[\protect\citeauthoryear{{Duc~et~al.}}{{Duc~et~al.}}{2011}]{Duc2011}
{Duc~et~al.} 2011, MNRAS, 417, 863

\bibitem[\protect\citeauthoryear{{Duquennoy} \& {Mayor}}{{Duquennoy} \&
  {Mayor}}{1991}]{Duquennoy1991}
{Duquennoy} A.,  {Mayor} M.,  1991, A\&A, 248, 485

\bibitem[\protect\citeauthoryear{{Elmegreen}, {Kaufman} \&
  {Thomasson}}{{Elmegreen} et~al.}{1993}]{Elmegreen1993}
{Elmegreen} B.~G.,  {Kaufman} M.,    {Thomasson} M.,  1993, ApJ, 412, 90

\bibitem[\protect\citeauthoryear{{Famaey} \& {Binney}}{{Famaey} \&
  {Binney}}{2005}]{Famaey2005}
{Famaey} B.,  {Binney} J.,  2005, MNRAS, 363, 603

\bibitem[\protect\citeauthoryear{{Famaey} \& {McGaugh}}{{Famaey} \&
  {McGaugh}}{2013}]{Famaey2013}
{Famaey} B.,  {McGaugh} S.,  2013, ArXiv e-prints

\bibitem[\protect\citeauthoryear{{Famaey} \& {McGaugh}}{{Famaey} \&
  {McGaugh}}{2012}]{Famaey2012}
{Famaey} B.,  {McGaugh} S.~S.,  2012, Living Reviews in Relativity, 15, 10

\bibitem[\protect\citeauthoryear{{Fellhauer} \& {Kroupa}}{{Fellhauer} \&
  {Kroupa}}{2006}]{Fellhauer2006}
{Fellhauer} M.,  {Kroupa} P.,  2006, MNRAS, 367, 1577

\bibitem[\protect\citeauthoryear{{Ferguson} \& {Binggeli}}{{Ferguson} \&
  {Binggeli}}{1994}]{Ferguson1994}
{Ferguson} H.~C.,  {Binggeli} B.,  1994, A\&ARv, 6, 67

\bibitem[\protect\citeauthoryear{{Ferreras}, {La Barbera}, {de la Rosa},
  {Vazdekis}, {de Carvalho}, {Falc{\'o}n-Barroso} \& {Ricciardelli}}{{Ferreras}
  et~al.}{2013}]{Ferreras2013}
{Ferreras} I.,  {La Barbera} F.,  {de la Rosa} I.~G.,  {Vazdekis} A.,  {de
  Carvalho} R.~R.,  {Falc{\'o}n-Barroso} J.,    {Ricciardelli} E.,  2013,
  MNRAS, 429, L15

\bibitem[\protect\citeauthoryear{{Forbes} \& {Kroupa}}{{Forbes} \&
  {Kroupa}}{2011}]{Forbes2011}
{Forbes} D.~A.,  {Kroupa} P.,  2011, PASA, 28, 77

\bibitem[\protect\citeauthoryear{{Fouquet}, {Hammer}, {Yang}, {Puech} \&
  {Flores}}{{Fouquet} et~al.}{2012}]{Fouquet2012}
{Fouquet} S.,  {Hammer} F.,  {Yang} Y.,  {Puech} M.,    {Flores} H.,  2012,
  MNRAS, 427, 1769

\bibitem[\protect\citeauthoryear{{Gentile}, {Famaey} \& {de Blok}}{{Gentile}
  et~al.}{2011}]{Gentile2011}
{Gentile} G.,  {Famaey} B.,    {de Blok} W.~J.~G.,  2011, A\&A, 527, A76

\bibitem[\protect\citeauthoryear{{Gieles}, {Baumgardt}, {Heggie} \&
  {Lamers}}{{Gieles} et~al.}{2010}]{Gieles2010}
{Gieles} M.,  {Baumgardt} H.,  {Heggie} D.~C.,    {Lamers} H.~J.~G.~L.~M.,
  2010, MNRAS, 408, L16

\bibitem[\protect\citeauthoryear{{Gilmore}, {Wilkinson}, {Wyse}, {Kleyna},
  {Koch}, {Evans} \& {Grebel}}{{Gilmore} et~al.}{2007}]{Gilmore2007}
{Gilmore} G.,  {Wilkinson} M.~I.,  {Wyse} R.~F.~G.,  {Kleyna} J.~T.,  {Koch}
  A.,  {Evans} N.~W.,    {Grebel} E.~K.,  2007, ApJ, 663, 948

\bibitem[\protect\citeauthoryear{{Governato}, {Zolotov}, {Pontzen},
  {Christensen}, {Oh}, {Brooks}, {Quinn}, {Shen} \& {Wadsley}}{{Governato}
  et~al.}{2012}]{Governato2012}
{Governato} F.,  {Zolotov} A.,  {Pontzen} A.,  {Christensen} C.,  {Oh} S.~H.,
  {Brooks} A.~M.,  {Quinn} T.,  {Shen} S.,    {Wadsley} J.,  2012, MNRAS, 422,
  1231

\bibitem[\protect\citeauthoryear{{Graham} \& {Colless}}{{Graham} \&
  {Colless}}{1997}]{Graham1997}
{Graham} A.,  {Colless} M.,  1997, MNRAS, 287, 221

\bibitem[\protect\citeauthoryear{{Guo}, {White}, {Boylan-Kolchin}, {De Lucia},
  {Kauffmann}, {Lemson}, {Li}, {Springel} \& {Weinmann}}{{Guo}
  et~al.}{2011}]{Guo2011}
{Guo} Q.,  {White} S.,  {Boylan-Kolchin} M.,  {De Lucia} G.,  {Kauffmann} G.,
  {Lemson} G.,  {Li} C.,  {Springel} V.,    {Weinmann} S.,  2011, MNRAS, 413,
  101

\bibitem[\protect\citeauthoryear{{Hammer}, {Yang}, {Fouquet}, {Pawlowski},
  {Kroupa}, {Puech}, {Flores} \& {Wang}}{{Hammer} et~al.}{2013}]{Hammer2013}
{Hammer} F.,  {Yang} Y.,  {Fouquet} S.,  {Pawlowski} M.~S.,  {Kroupa} P.,
  {Puech} M.,  {Flores} H.,    {Wang} J.,  2013, MNRAS

\bibitem[\protect\citeauthoryear{{Hargreaves}, {Gilmore} \&
  {Annan}}{{Hargreaves} et~al.}{1996}]{Hargreaves1996}
{Hargreaves} J.~C.,  {Gilmore} G.,    {Annan} J.~D.,  1996, MNRAS, 279, 108

\bibitem[\protect\citeauthoryear{{Hees}, {Famaey}, {Angus} \& {Gentile}}{{Hees}
  et~al.}{2016}]{Hees2016}
{Hees} A.,  {Famaey} B.,  {Angus} G.~W.,    {Gentile} G.,  2016, MNRAS, 455,
  449

\bibitem[\protect\citeauthoryear{{Ibata}, {Ibata}, {Famaey} \& {Lewis}}{{Ibata}
  et~al.}{2014}]{Ibata2014}
{Ibata} N.~G.,  {Ibata} R.~A.,  {Famaey} B.,    {Lewis} G.~F.,  2014, Nature,
  511, 563

\bibitem[\protect\citeauthoryear{{Ibata~et~al.}}{{Ibata~et~al.}}{2013}]{Ibata2013}
{Ibata~et~al.} 2013, Nature, 493, 62

\bibitem[\protect\citeauthoryear{{Janz}, {Cappellari}, {Romanowsky}, {Ciotti},
  {Alabi} \& {Forbes}}{{Janz} et~al.}{2016}]{Janz2016}
{Janz} J.,  {Cappellari} M.,  {Romanowsky} A.~J.,  {Ciotti} L.,  {Alabi} A.,
  {Forbes} D.~A.,  2016, MNRAS, 461, 2367

\bibitem[\protect\citeauthoryear{{Kalirai}, {Hansen}, {Kelson}, {Reitzel},
  {Rich} \& {Richer}}{{Kalirai} et~al.}{2008}]{Kalirai2008}
{Kalirai} J.~S.,  {Hansen} B.~M.~S.,  {Kelson} D.~D.,  {Reitzel} D.~B.,  {Rich}
  R.~M.,    {Richer} H.~B.,  2008, ApJ, 676, 594

\bibitem[\protect\citeauthoryear{{Kouwenhoven} \& {de Grijs}}{{Kouwenhoven} \&
  {de Grijs}}{2008}]{Kouwenhoven2008}
{Kouwenhoven} M.~B.~N.,  {de Grijs} R.,  2008, A\&A, 480, 103

\bibitem[\protect\citeauthoryear{{Kroupa}}{{Kroupa}}{1997}]{Kroupa1997}
{Kroupa} P.,  1997, New Astronomy, 2, 139

\bibitem[\protect\citeauthoryear{{Kroupa}}{{Kroupa}}{2001}]{Kroupa2001}
{Kroupa} P.,  2001, MNRAS, 322, 231

\bibitem[\protect\citeauthoryear{{Kroupa}}{{Kroupa}}{2012}]{Kroupa2012}
{Kroupa} P.,  2012, PASA, 29, 395

\bibitem[\protect\citeauthoryear{{Kroupa}}{{Kroupa}}{2015}]{Kroupa2015}
{Kroupa} P.,  2015, Canadian Journal of Physics, 93, 169

\bibitem[\protect\citeauthoryear{{Kroupa}, {Theis} \& {Boily}}{{Kroupa}
  et~al.}{2005}]{Kroupa2005}
{Kroupa} P.,  {Theis} C.,    {Boily} C.~M.,  2005, A\&A, 431, 517

\bibitem[\protect\citeauthoryear{{Kroupa}, {Weidner}, {Pflamm-Altenburg},
  {Thies}, {Dabringhausen}, {Marks} \& {Maschberger}}{{Kroupa}
  et~al.}{2013}]{Kroupa2013}
{Kroupa} P.,  {Weidner} C.,  {Pflamm-Altenburg} J.,  {Thies} I.,
  {Dabringhausen} J.,  {Marks} M.,    {Maschberger} T.,  2013, {The Stellar and
  Sub-Stellar Initial Mass Function of Simple and Composite Populations}.
p.~115

\bibitem[\protect\citeauthoryear{{Kroupa~et~al.}}{{Kroupa~et~al.}}{2010}]{Kroupa2010}
{Kroupa~et~al.} 2010, A\&A, 523, 32

\bibitem[\protect\citeauthoryear{{Leier}, {Ferreras}, {Saha}, {Charlot},
  {Bruzual} \& {La Barbera}}{{Leier} et~al.}{2016}]{Leier2016}
{Leier} D.,  {Ferreras} I.,  {Saha} P.,  {Charlot} S.,  {Bruzual} G.,    {La
  Barbera} F.,  2016, MNRAS

\bibitem[\protect\citeauthoryear{{Li}, {De Lucia} \& {Helmi}}{{Li}
  et~al.}{2010}]{Li2010}
{Li} Y.-S.,  {De Lucia} G.,    {Helmi} A.,  2010, MNRAS, 401, 2036

\bibitem[\protect\citeauthoryear{{Lisker}}{{Lisker}}{2009}]{Lisker2009}
{Lisker} T.,  2009, Astronomische Nachrichten, 330, 1043

\bibitem[\protect\citeauthoryear{{L{\'o}pez-Corredoira} \&
  {Kroupa}}{{L{\'o}pez-Corredoira} \& {Kroupa}}{2016}]{LopezCorredoira2016}
{L{\'o}pez-Corredoira} M.,  {Kroupa} P.,  2016, ApJ, 817, 75

\bibitem[\protect\citeauthoryear{{L{\"u}ghausen}, {Famaey} \&
  {Kroupa}}{{L{\"u}ghausen} et~al.}{2014}]{Lueghausen2014}
{L{\"u}ghausen} F.,  {Famaey} B.,    {Kroupa} P.,  2014, MNRAS, 441, 2497

\bibitem[\protect\citeauthoryear{{L{\"u}ghausen}, {Famaey} \&
  {Kroupa}}{{L{\"u}ghausen} et~al.}{2015}]{Lueghausen2015}
{L{\"u}ghausen} F.,  {Famaey} B.,    {Kroupa} P.,  2015, Canadian Journal of
  Physics, 93, 232

\bibitem[\protect\citeauthoryear{{Marks} \& {Kroupa}}{{Marks} \&
  {Kroupa}}{2011}]{Marks2011}
{Marks} M.,  {Kroupa} P.,  2011, MNRAS, 417, 1702

\bibitem[\protect\citeauthoryear{{Marks}, {Kroupa}, {Dabringhausen} \&
  {Pawlowski}}{{Marks} et~al.}{2012}]{Marks2012b}
{Marks} M.,  {Kroupa} P.,  {Dabringhausen} J.,    {Pawlowski} M.~S.,  2012,
  MNRAS, 422, 2246

\bibitem[\protect\citeauthoryear{{Mateo}}{{Mateo}}{1998}]{Mateo1998}
{Mateo} M.~L.,  1998, ARA\&A, 36, 435

\bibitem[\protect\citeauthoryear{{McConnachie} \& {C{\^o}t{\'e}}}{{McConnachie}
  \& {C{\^o}t{\'e}}}{2010}]{McConnachie2010}
{McConnachie} A.~W.,  {C{\^o}t{\'e}} P.,  2010, ApJ, 722, L209

\bibitem[\protect\citeauthoryear{{McGaugh} \& {Milgrom}}{{McGaugh} \&
  {Milgrom}}{2013}]{McGaugh2013b}
{McGaugh} S.,  {Milgrom} M.,  2013, ApJ, 775, 139

\bibitem[\protect\citeauthoryear{{McGaugh} \& {Wolf}}{{McGaugh} \&
  {Wolf}}{2010}]{McGaugh2010}
{McGaugh} S.~S.,  {Wolf} J.,  2010, ApJ, 722, 248

\bibitem[\protect\citeauthoryear{{Metz} \& {Kroupa}}{{Metz} \&
  {Kroupa}}{2007}]{Metz2007}
{Metz} M.,  {Kroupa} P.,  2007, MNRAS, 376, 387

\bibitem[\protect\citeauthoryear{{Metz}, {Kroupa} \& {Libeskind}}{{Metz}
  et~al.}{2008}]{Metz2008}
{Metz} M.,  {Kroupa} P.,    {Libeskind} N.~I.,  2008, ApJ, 680, 287

\bibitem[\protect\citeauthoryear{{Milgrom}}{{Milgrom}}{1983a}]{Milgrom1983b}
{Milgrom} M.,  1983a, ApJ, 270, 371

\bibitem[\protect\citeauthoryear{{Milgrom}}{{Milgrom}}{1983b}]{Milgrom1983a}
{Milgrom} M.,  1983b, ApJ, 270, 365

\bibitem[\protect\citeauthoryear{{Milgrom}}{{Milgrom}}{2010}]{Milgrom2010}
{Milgrom} M.,  2010, MNRAS, 403, 886

\bibitem[\protect\citeauthoryear{{Milgrom}}{{Milgrom}}{2012}]{Milgrom2012}
{Milgrom} M.,  2012, Physical Review Letters, 109, 131101

\bibitem[\protect\citeauthoryear{{Milgrom} \& {Sanders}}{{Milgrom} \&
  {Sanders}}{2003}]{Milgrom2003}
{Milgrom} M.,  {Sanders} R.~H.,  2003, ApJ, 599, L25

\bibitem[\protect\citeauthoryear{{Minor}}{{Minor}}{2013}]{Minor2013}
{Minor} Q.~E.,  2013, ApJ, 779, 116

\bibitem[\protect\citeauthoryear{{Mirabel}, {Dottori} \& {Lutz}}{{Mirabel}
  et~al.}{1992}]{Mirabel1992}
{Mirabel} I.~F.,  {Dottori} H.,    {Lutz} D.,  1992, A\&A, 256, L19

\bibitem[\protect\citeauthoryear{{Misgeld} \& {Hilker}}{{Misgeld} \&
  {Hilker}}{2011}]{Misgeld2011}
{Misgeld} I.,  {Hilker} M.,  2011, MNRAS, 414, 3699

\bibitem[\protect\citeauthoryear{{Monreal-Ibero}, {Colina}, {Arribas} \&
  {Garc{\'{\i}}a-Mar{\'{\i}}n}}{{Monreal-Ibero} et~al.}{2007}]{Monreal2007}
{Monreal-Ibero} A.,  {Colina} L.,  {Arribas} S.,
  {Garc{\'{\i}}a-Mar{\'{\i}}n} M.,  2007, A\&A, 472, 421

\bibitem[\protect\citeauthoryear{{Mu{\~n}oz}, {Geha} \& {Willman}}{{Mu{\~n}oz}
  et~al.}{2010}]{Munoz2010}
{Mu{\~n}oz} R.~R.,  {Geha} M.,    {Willman} B.,  2010, AJ, 140, 138

\bibitem[\protect\citeauthoryear{{O{\~n}orbe}, {Boylan-Kolchin}, {Bullock},
  {Hopkins}, {Kere{\v s}}, {Faucher-Gigu{\`e}re}, {Quataert} \&
  {Murray}}{{O{\~n}orbe} et~al.}{2015}]{Onorbe2015}
{O{\~n}orbe} J.,  {Boylan-Kolchin} M.,  {Bullock} J.~S.,  {Hopkins} P.~F.,
  {Kere{\v s}} D.,  {Faucher-Gigu{\`e}re} C.-A.,  {Quataert} E.,    {Murray}
  N.,  2015, MNRAS, 454, 2092

\bibitem[\protect\citeauthoryear{{Okazaki} \& {Taniguchi}}{{Okazaki} \&
  {Taniguchi}}{2000}]{Okazaki2000}
{Okazaki} T.,  {Taniguchi} Y.,  2000, ApJ, 543, 149

\bibitem[\protect\citeauthoryear{{Pawlowski}, {Kroupa}, {Angus}, {de Boer},
  {Famaey} \& {Hensler}}{{Pawlowski} et~al.}{2012}]{Pawlowski2012b}
{Pawlowski} M.~S.,  {Kroupa} P.,  {Angus} G.,  {de Boer} K.~S.,  {Famaey} B.,
   {Hensler} G.,  2012, MNRAS, 424, 80

\bibitem[\protect\citeauthoryear{{Pawlowski}, {Kroupa} \& {Jerjen}}{{Pawlowski}
  et~al.}{2013}]{Pawlowski2013}
{Pawlowski} M.~S.,  {Kroupa} P.,    {Jerjen} H.,  2013, MNRAS, 435, 1928

\bibitem[\protect\citeauthoryear{{Pawlowski}, {Pflamm-Altenburg} \&
  {Kroupa}}{{Pawlowski} et~al.}{2012}]{Pawlowski2012a}
{Pawlowski} M.~S.,  {Pflamm-Altenburg} J.,    {Kroupa} P.,  2012, MNRAS,
  p.~2990

\bibitem[\protect\citeauthoryear{{Pawlowski~et~al.}}{{Pawlowski~et~al.}}{2014}]{Pawlowski2014}
{Pawlowski~et~al.} 2014, MNRAS, 442, 2362

\bibitem[\protect\citeauthoryear{{Peacock}, {Zepf}, {Maccarone}, {Kundu},
  {Gonzalez}, {Lehmer} \& {Maraston}}{{Peacock} et~al.}{2014}]{Peacock2014}
{Peacock} M.~B.,  {Zepf} S.~E.,  {Maccarone} T.~J.,  {Kundu} A.,  {Gonzalez}
  A.~H.,  {Lehmer} B.~D.,    {Maraston} C.,  2014, ApJ, 784, 162

\bibitem[\protect\citeauthoryear{{Ploeckinger}, {Hensler}, {Recchi}, {Mitchell}
  \& {Kroupa}}{{Ploeckinger} et~al.}{2014}]{Ploeckinger2014}
{Ploeckinger} S.,  {Hensler} G.,  {Recchi} S.,  {Mitchell} N.,    {Kroupa} P.,
  2014, MNRAS, 437, 3980

\bibitem[\protect\citeauthoryear{{Ploeckinger}, {Recchi}, {Hensler} \&
  {Kroupa}}{{Ploeckinger} et~al.}{2015}]{Ploeckinger2015}
{Ploeckinger} S.,  {Recchi} S.,  {Hensler} G.,    {Kroupa} P.,  2015, MNRAS,
  447, 2512

\bibitem[\protect\citeauthoryear{{Recchi}, {Theis}, {Kroupa} \&
  {Hensler}}{{Recchi} et~al.}{2007}]{Recchi2007}
{Recchi} S.,  {Theis} C.,  {Kroupa} P.,    {Hensler} G.,  2007, A\&A, 470, L5

\bibitem[\protect\citeauthoryear{{Richtler}, {Famaey}, {Gentile} \&
  {Schuberth}}{{Richtler} et~al.}{2011}]{Richtler2011}
{Richtler} T.,  {Famaey} B.,  {Gentile} G.,    {Schuberth} Y.,  2011, A\&A,
  531, A100

\bibitem[\protect\citeauthoryear{{Richtler}, {Schuberth}, {Hilker}, {Dirsch},
  {Bassino} \& {Romanowsky}}{{Richtler} et~al.}{2008}]{Richtler2008}
{Richtler} T.,  {Schuberth} Y.,  {Hilker} M.,  {Dirsch} B.,  {Bassino} L.,
  {Romanowsky} A.~J.,  2008, A\&A, 478, L23

\bibitem[\protect\citeauthoryear{{Samurovi{\'c}}}{{Samurovi{\'c}}}{2010}]{Samurovic2010}
{Samurovi{\'c}} S.,  2010, A\&A, 514, A95

\bibitem[\protect\citeauthoryear{{Samurovi{\'c}}}{{Samurovi{\'c}}}{2014}]{Samurovic2014}
{Samurovi{\'c}} S.,  2014, A\&A, 570, A132

\bibitem[\protect\citeauthoryear{{Samurovi{\'c}}}{{Samurovi{\'c}}}{2016}]{Samurovic2016}
{Samurovi{\'c}} S.,  2016, Ap\&SS, 361, 199

\bibitem[\protect\citeauthoryear{{Sanders} \& {McGaugh}}{{Sanders} \&
  {McGaugh}}{2002}]{Sanders2002}
{Sanders} R.~H.,  {McGaugh} S.~S.,  2002, ARA\&A, 40, 263

\bibitem[\protect\citeauthoryear{{Schuberth}, {Richtler}, {Dirsch}, {Hilker},
  {Larsen}, {Kissler-Patig} \& {Mebold}}{{Schuberth}
  et~al.}{2006}]{Schuberth2006}
{Schuberth} Y.,  {Richtler} T.,  {Dirsch} B.,  {Hilker} M.,  {Larsen} S.~S.,
  {Kissler-Patig} M.,    {Mebold} U.,  2006, A\&A, 459, 391

\bibitem[\protect\citeauthoryear{{Simon~et~al.}}{{Simon~et~al.}}{2011}]{Simon2011}
{Simon~et~al.} 2011, ApJ, 733, 46

\bibitem[\protect\citeauthoryear{{Simonneau} \& {Prada}}{{Simonneau} \&
  {Prada}}{2004}]{Simonneau2004}
{Simonneau} E.,  {Prada} F.,  2004, Revista Mexicana de Astronom\'{i}a y
  Astrof\'{i}sica, 40, 69

\bibitem[\protect\citeauthoryear{{Strigari}, {Bullock}, {Kaplinghat}, {Simon},
  {Geha}, {Willman} \& {Walker}}{{Strigari} et~al.}{2008}]{Strigari2008}
{Strigari} L.~E.,  {Bullock} J.~S.,  {Kaplinghat} M.,  {Simon} J.~D.,  {Geha}
  M.,  {Willman} B.,    {Walker} M.~G.,  2008, Nature, 454, 1096

\bibitem[\protect\citeauthoryear{{Teyssier}}{{Teyssier}}{2002}]{Teyssier2002}
{Teyssier} R.,  2002, A\&A, 385, 337

\bibitem[\protect\citeauthoryear{{Thomas}, {Maraston}, {Bender} \& {Mendes de
  Oliveira}}{{Thomas} et~al.}{2005}]{Thomas2005}
{Thomas} D.,  {Maraston} C.,  {Bender} R.,    {Mendes de Oliveira} C.,  2005,
  ApJ, 621, 673

\bibitem[\protect\citeauthoryear{{Tiret}, {Combes}, {Angus}, {Famaey} \&
  {Zhao}}{{Tiret} et~al.}{2007}]{Tiret2007}
{Tiret} O.,  {Combes} F.,  {Angus} G.~W.,  {Famaey} B.,    {Zhao} H.~S.,  2007,
  A\&A, 476, L1

\bibitem[\protect\citeauthoryear{{Tollerud~et~al.}}{{Tollerud~et~al.}}{2012}]{Tollerud2012}
{Tollerud~et~al.} 2012, ApJ, 752, 45

\bibitem[\protect\citeauthoryear{{Toloba~et~al.}}{{Toloba~et~al.}}{2014}]{Toloba2014}
{Toloba~et~al.} 2014, ApJS, 215, 17

\bibitem[\protect\citeauthoryear{{Tortora}, {Romanowsky}, {Cardone},
  {Napolitano} \& {Jetzer}}{{Tortora} et~al.}{2014}]{Tortora2014}
{Tortora} C.,  {Romanowsky} A.~J.,  {Cardone} V.~F.,  {Napolitano} N.~R.,
  {Jetzer} P.,  2014, MNRAS, 438, L46

\bibitem[\protect\citeauthoryear{{van Dokkum} \& {Conroy}}{{van Dokkum} \&
  {Conroy}}{2010}]{vanDokkum2010}
{van Dokkum} P.~G.,  {Conroy} C.,  2010, Nature, 468, 940

\bibitem[\protect\citeauthoryear{{Vudragovi\'{c}}, {Samurovi\'{c}} \&
  {Jovanovi\'{c}}}{{Vudragovi\'{c}} et~al.}{2016}]{Vudragovic2016}
{Vudragovi\'{c}} A.,  {Samurovi\'{c}} S.,    {Jovanovi\'{c}} M.,  2016, A\&A

\bibitem[\protect\citeauthoryear{{Walker}, {Belokurov}, {Evans}, {Irwin},
  {Mateo}, {Olszewski} \& {Gilmore}}{{Walker} et~al.}{2009}]{Walker2009}
{Walker} M.~G.,  {Belokurov} V.,  {Evans} N.~W.,  {Irwin} M.~J.,  {Mateo} M.,
  {Olszewski} E.~W.,    {Gilmore} G.,  2009, ApJ, 694, L144

\bibitem[\protect\citeauthoryear{{Weidner}, {Kroupa} \& {Larsen}}{{Weidner}
  et~al.}{2004}]{Weidner2004b}
{Weidner} C.,  {Kroupa} P.,    {Larsen} S.~S.,  2004, MNRAS, 350, 1503

\bibitem[\protect\citeauthoryear{{Weidner}, {Kroupa}, {Pflamm-Altenburg} \&
  {Vazdekis}}{{Weidner} et~al.}{2013}]{Weidner2013}
{Weidner} C.,  {Kroupa} P.,  {Pflamm-Altenburg} J.,    {Vazdekis} A.,  2013,
  MNRAS, 436, 3309

\bibitem[\protect\citeauthoryear{{Weisz}, {Dolphin}, {Skillman}, {Holtzman},
  {Gilbert}, {Dalcanton} \& {Williams}}{{Weisz} et~al.}{2014}]{Weisz2014}
{Weisz} D.~R.,  {Dolphin} A.~E.,  {Skillman} E.~D.,  {Holtzman} J.,  {Gilbert}
  K.~M.,  {Dalcanton} J.~J.,    {Williams} B.~F.,  2014, ApJ, 789, 148

\bibitem[\protect\citeauthoryear{{Wetzstein}, {Naab} \& {Burkert}}{{Wetzstein}
  et~al.}{2007}]{Wetzstein2007}
{Wetzstein} M.,  {Naab} T.,    {Burkert} A.,  2007, MNRAS, 375, 805

\bibitem[\protect\citeauthoryear{{Willman}, {Geha}, {Strader}, {Strigari},
  {Simon}, {Kirby}, {Ho} \& {Warres}}{{Willman} et~al.}{2011}]{Willman2011}
{Willman} B.,  {Geha} M.,  {Strader} J.,  {Strigari} L.~E.,  {Simon} J.~D.,
  {Kirby} E.,  {Ho} N.,    {Warres} A.,  2011, AJ, 142, 128

\bibitem[\protect\citeauthoryear{{Wolf}, {Martinez}, {Bullock}, {Kaplinghat},
  {Geha}, {Mu{\~n}oz}, {Simon} \& {Avedo}}{{Wolf} et~al.}{2010}]{Wolf2010}
{Wolf} J.,  {Martinez} G.~D.,  {Bullock} J.~S.,  {Kaplinghat} M.,  {Geha} M.,
  {Mu{\~n}oz} R.~R.,  {Simon} J.~D.,    {Avedo} F.~F.,  2010, MNRAS, 406, 1220

\bibitem[\protect\citeauthoryear{{Yang}, {Hammer}, {Fouquet}, {Flores},
  {Puech}, {Pawlowski} \& {Kroupa}}{{Yang} et~al.}{2014}]{Yang2014}
{Yang} Y.,  {Hammer} F.,  {Fouquet} S.,  {Flores} H.,  {Puech} M.,  {Pawlowski}
  M.~S.,    {Kroupa} P.,  2014, MNRAS, 442, 2419

\bibitem[\protect\citeauthoryear{{Yoshida}, {Yagi}, {Komiyama}, {Furusawa},
  {Kashikawa}, {Koyama}, {Yamanoi}, {Hattori} \& {Okamura}}{{Yoshida}
  et~al.}{2008}]{Yoshida2008}
{Yoshida} M.,  {Yagi} M.,  {Komiyama} Y.,  {Furusawa} H.,  {Kashikawa} N.,
  {Koyama} Y.,  {Yamanoi} H.,  {Hattori} T.,    {Okamura} S.,  2008, ApJ, 688,
  918

\bibitem[\protect\citeauthoryear{{Young~et~al.}}{{Young~et~al.}}{2011}]{Young2011}
{Young~et~al.} 2011, MNRAS, 414, 940

\bibitem[\protect\citeauthoryear{{Zhao} \& {Famaey}}{{Zhao} \&
  {Famaey}}{2006}]{Zhao2006}
{Zhao} H.~S.,  {Famaey} B.,  2006, ApJ, 638, L9

\bibitem[\protect\citeauthoryear{{Zwicky}}{{Zwicky}}{1956}]{Zwicky1956}
{Zwicky} F.,  1956, Ergebnisse der exakten Naturwissenschaften, 29, 344

\end{thebibliography}

\label{lastpage}

\end{document}